\documentclass[a4paper,fleqn,usenatbib]{mnras}

\usepackage{newtxtext,newtxmath}
\usepackage[T1]{fontenc}
\usepackage{ae,aecompl}

\usepackage{graphicx}	% Including figure files
\usepackage{amsmath}	% Advanced maths commands
\usepackage{amssymb}	% Extra maths symbols
\usepackage{ulem}       % strikethrough function
\usepackage[dvipsnames]{xcolor}

\newcommand{\Rs}{R_{\star}}

\newcommand{\aRs}{a/\Rs}
\newcommand{\Tmid}{T_{\textnormal{mid}}}

\newcommand{\um}{\mu\textnormal{m}}

\newcommand{\Cabun}{[\textnormal{C}/\textnormal{H}]}
\newcommand{\Oabun}{[\textnormal{O}/\textnormal{H}]}
\newcommand{\kir}{\kappa_\textnormal{IR}}
\newcommand{\Z}{[\textnormal{M}/\textnormal{H}]}
\title[WASP-121b emission spectrum across $0.8$--$1.1\,\um$]{An emission spectrum for WASP-121b measured across the $0.8$--$1.1\,\um$ wavelength range using the Hubble Space Telescope}

\author[T.\ Mikal-Evans et al.]{
Thomas Mikal-Evans,$^{1}$\thanks{E-mail: tmevans@mit.edu}
David K.\ Sing,$^{2,3}$ Jayesh M.\ Goyal,$^{4}$ Benjamin Drummond,$^{4}$
\newauthor Aarynn L.\ Carter,$^{4}$ Gregory W.\ Henry,$^{5}$ Hannah R.\ Wakeford,$^{6}$ Nikole K.\ Lewis,$^{7}$
\newauthor Mark S.\ Marley,$^{8}$ Pascal Tremblin,$^{9}$ Nikolay Nikolov,$^{3}$ Tiffany Kataria,$^{10}$
\newauthor Drake Deming,$^{11}$ Gilda E.\ Ballester$^{12}$ 
\\
$^{1}$Kavli Institute for Astrophysics and Space Research, Massachusetts Institute of Technology, 77 Massachusetts Avenue,\\
Cambridge, MA 02139, USA\\
$^{2}$Department of Earth and Planetary Sciences, Johns Hopkins University, Baltimore, MD, USA\\
$^{3}$Department of Physics \& Astronomy, Johns Hopkins University, Baltimore, MD, USA \\
$^{4}$Physics and Astronomy, Stocker Road, University of Exeter, Exeter, EX4 3RF, UK\\
$^{5}$Center of Excellence in Information Systems, Tennessee State University, Nashville, TN 37209, USA\\
$^{6}$Space Telescope Science Institute, 3700 San Martin Drive, Baltimore, Maryland 21218, USA \\
$^{7}$Department of Astronomy and Carl Sagan Institute, Cornell University, 122 Sciences Drive, 14853, Ithaca, NY, USA\\
$^{8}$NASA Ames Research Center, Moffett Field, California, USA\\
$^{9}$Maison de la Simulation, CEA, CNRS, Univ\ Paris-Sud, UVSQ, Universit\'{e} Paris-Saclay, F-91191 Gif-sur-Yvette, France\\
$^{10}$NASA Jet Propulsion Laboratory, 4800 Oak Grove Drive, Pasadena, CA 91109, USA\\
$^{11}$Department of Astronomy, University of Maryland at College Park, College Park, MD 20742, USA\\
$^{12}$Lunar and Planetary Laboratory, University of Arizona, Tucson, AZ 85721, USA
}

\date{Accepted 2019 June 14. Received 2019 June 4; in original form 2019 March 28}

\pubyear{2019}

\begin{document}
\label{firstpage}
\pagerange{\pageref{firstpage}--\pageref{lastpage}}
\maketitle

\begin{abstract}
WASP-121b is a transiting gas giant exoplanet orbiting close to its Roche limit, with an inflated radius nearly double that of Jupiter and a dayside temperature comparable to a late M dwarf photosphere. Secondary eclipse observations covering the $1.1$-$1.6\,\um$ wavelength range have revealed an atmospheric thermal inversion on the dayside hemisphere, likely caused by high altitude absorption at optical wavelengths. Here we present secondary eclipse observations made with the \textit{Hubble Space Telescope} Wide Field Camera 3 spectrograph that extend the wavelength coverage from $1.1\,\um$ down to $0.8\,\um$. To determine the atmospheric properties from the measured eclipse spectrum, we performed a retrieval analysis assuming chemical equilibrium, with the effects of thermal dissociation and ionization included. Our best-fit model provides a good fit to the data with reduced $\chi^2_\nu=1.04$. The data diverge from a blackbody spectrum and instead exhibit emission due to H$^-$ shortward of $1.1\,\um$. The best-fit model does not reproduce a previously reported bump in the spectrum at $1.25\um$, possibly indicating this feature is a statistical fluctuation in the data rather than a VO emission band as had been tentatively suggested. We estimate an atmospheric metallicity of $\Z = {1.09}_{-0.69}^{+0.57}$, and fit for the carbon and oxygen abundances separately, obtaining $\Cabun = {-0.29}_{-0.48}^{+0.61}$ and $\Oabun = {0.18}_{-0.60}^{+0.64}$. The corresponding carbon-to-oxygen ratio is $\textnormal{C/O} = 0.49_{-0.37}^{+0.65}$, which encompasses the solar value of 0.54, but has a large uncertainty. 
\end{abstract}

\begin{keywords}
planets and satellites: atmospheres -- planets and satellites: gaseous planets -- techniques: spectroscopic
\end{keywords}

%%%%%%%%%%%%%%%%%%%%%%%%%%%%%%%%%%%%%%%%%%%%%%%%%%

%%%%%%%%%%%%%%%%% BODY OF PAPER %%%%%%%%%%%%%%%%%%
\section{Introduction}

Eclipse observations made as an exoplanet passes through superior conjunction allow the emission from the planetary dayside hemisphere to be inferred. Dozens of eclipse measurements have been published for broad photometric passbands, most notably from the \textit{Kepler} space mission \citep[e.g.][]{2009Sci...325..709B,2011MNRAS.417L..88K,2013ApJ...771...26F,2015PASP..127.1113A,2015ApJ...804..150E}, the \textit{Spitzer Space Telescope} Infrared Array Camera (IRAC) \citep[e.g.][]{2005ApJ...626..523C,2010ApJ...720.1569K,2010Natur.464.1161S,2014ApJ...781..116B,2019arXiv190107040G}, Infrared Spectrograph (IRS) \citep{2012ApJ...754..136S,2014ApJ...797...42C}, and Multiband Imaging Photometer \citep{2005Natur.434..740D,2008ApJ...686.1341C,2012ApJ...752...81C}, as well as ground-based telescopes \citep[e.g.][]{2009A&A...493L..31S,2010MNRAS.404L.114G,2011A&A...528A..49D,2019A&A...624A..62M,2019A&A...625A..80K}. A smaller number of spectroscopic eclipse measurements have also been published, including two with the \textit{Spitzer} IRS \citep{2007Natur.445..892R,2007ApJ...658L.115G,2008Natur.456..767G}, two with the \textit{Hubble Space Telescope} (HST) Near-Infrared Camera Multi-Object Spectrometer \citep{2009ApJ...690L.114S,2009ApJ...704.1616S}, and fourteen with the HST Wide Field Camera 3 (WFC3) infrared spectrograph \citep{2014ApJ...783..113W,2014ApJ...785..148R,2014Sci...346..838S,2014ApJ...791...36S,2014ApJ...793L..27K,2018AJ....156...17K,2014ApJ...795..166C,2015ApJ...806..146H,2016AJ....152..203L,2017AJ....154..158B,2017ApJ...850L..32S,2017Natur.548...58E,2018MNRAS.474.1705N,2018ApJ...855L..30A,2018AJ....156...10M}. Spectroscopic observations are of particular value by allowing opacity bands to be resolved, which in turn encode information about chemical abundances and the vertical temperature profile of the atmosphere.

The WFC3 instrument offers two grisms for infrared spectroscopy:\ G102 covering wavelengths $0.8$-$1.1\,\um$ and G141 covering wavelengths $1.1$-$1.6\,\um$. All of the exoplanet emission spectra published to date have used the G141 grism for two main reasons: (i) the longer wavelength coverage provides a more favorable planet-to-star brightness ratio; and (ii) the G141 passband provides access to stronger opacity bands than the G102 passband, in particular the $1.4\,\um$ H$_2$O band. The observations made with WFC3 G141 have resulted in detections of H$_2$O absorption for WASP-43b \citep{2014Sci...346..838S,2014ApJ...793L..27K}, HD\,189733b \citep{2014ApJ...795..166C}, HD\,209458b \citep{2016AJ....152..203L}, and Kepler-13Ab \citep{2017AJ....154..158B}, and H$_2$O emission for WASP-121b \citep{2017Natur.548...58E}. Other spectral features reported include TiO emission for WASP-33b \citep{2015ApJ...806..146H} and CO absorption for WASP-18b \citep{2017ApJ...850L..32S}. The remaining seven spectra do not exhibit significant spectral features and typically appear blackbody-like \citep{2014ApJ...783..113W,2014ApJ...785..148R,2018MNRAS.474.1705N,2018AJ....156...10M,2018AJ....156...17K}, while the CO absorption feature claimed for WASP-18b has been challenged \citep{2018ApJ...855L..30A}. Recent studies \citep{2018ApJ...855L..30A,2018ApJ...866...27L,2018A&A...617A.110P,2018AJ....156...17K} have suggested that continuum opacity due to H$^-$ and thermal dissociation of H$_2$O itself can explain the lack of spectral features observed for the ultrahot Jupiters, such as WASP-18b, HAT-P-7b, and WASP-103b, which have dayside temperatures well in excess of 2000\,K. 

Another such ultrahot Jupiter, WASP-121b, is the subject of this study. Discovered by \cite{2016MNRAS.tmp..312D}, WASP-121b has an exceptionally inflated radius ($1.8\,R_J$) and a dayside photospheric temperature of approximately 2700\,K \citep{2017Natur.548...58E}. The high temperature results from WASP-121b orbiting at a distance of only 0.025\,AU from its F6V host star, where it is subjected to strong tidal forces and may be undergoing atmospheric escape via Roche lobe overflow \citep{2016MNRAS.tmp..312D}. Observational support for this picture has recently been provided by near-ultraviolet (NUV) transit measurements made with \textit{Swift} UVOT that are significantly deeper than those measured at optical wavelengths \citep{2019A&A...623A..57S}. This could be explained by an extended atmosphere filling the Roche lobe, which is relatively opaque at NUV wavelengths due, for instance, to heavy metal absorption lines.

At near-infrared wavelengths, the HST WFC3 spectrograph has been used to measure both the transmission spectrum \citep{2016ApJ...822L...4E} and emission spectrum \citep{2017Natur.548...58E} of WASP-121b. Absorption due to the $1.4\,\um$ H$_2$O band is observed in the transmission spectrum, while this same band is seen in emission at secondary eclipse, revealing a thermal inversion on the dayside hemisphere. The latter indicates significant absorption of incident stellar radiation at NUV-optical wavelengths for pressures less than $\sim 100$\,mbar on the dayside hemisphere \citep[e.g.][]{2010A&A...520A..27G}. This is consistent with the low geometric albedo of $A_g = 0.16 \pm 0.11$ inferred by \cite{2019A&A...624A..62M} from the $z^\prime$ secondary eclipse measurement of \cite{2016MNRAS.tmp..312D}. Possible absorbers include TiO and VO, both of which have strong opacity bands throughout the optical \citep{2003ApJ...594.1011H,2008ApJ...678.1419F}. Indeed, the optical transmission spectrum measured using the HST Space Telescope Imaging Spectrograph (STIS) does show evidence for VO absorption at the day-night limb, although TiO is not seen \citep{2018AJ....156..283E}. A steep rise toward NUV wavelengths is also recovered in the STIS transmission spectrum, which may be caused by the same absorber/s responsible for the deep \textit{Swift} UVOT transit. One candidate proposed in \cite{2018AJ....156..283E} is SH, which has been predicted as a product of non-equilibrium chemistry in hot Jupiter atmospheres by \cite{2009ApJ...701L..20Z}, and, if present on the dayside hemisphere, could potentially produce the thermal inversion. As alluded to above, absorption by heavy metals such as Fe and Mg might also simultaneously explain the deep NUV transits and dayside thermal inversion. At optical wavelengths, other candidate absorbers that could play a role in generating the thermal inversion include H$^-$ ions and molecules such as NaH, MgH, FeH, SiO, AlO, and CaO \citep{2018ApJ...866...27L,2018A&A...617A.110P,2019MNRAS.485.5817G}, although no compelling evidence has been claimed for any of these species based on the published transmission spectrum \citep{2016ApJ...822L...4E,2018AJ....156..283E}.

Unlike the transmission spectrum --- which probes a region of the atmosphere very different to the ultrahot dayside of WASP-121b --- a detection of one or more strong optical absorbers in the emission spectrum would provide a definitive link between the radiatively active species present and the thermal inversion. Motivated by this, we acquired secondary eclipse observations of WASP-121b with the G102 grism of WFC3, extending the wavelength coverage into the red optical where emission bands due to species such as TiO, VO, and FeH may be detectable, as well as H$^-$ continuum opacity. We describe these observations and our data reduction in Section \ref{sec:observations_datared}. Our analyses of the white and spectroscopic light curves are presented in Sections \ref{sec:whitelc} and \ref{sec:speclcs}, respectively. The results are discussed in Section \ref{sec:discussion} and we conclude in Section \ref{sec:conclusion}. 

\begin{figure*}
\centering  % this centres figure in column
\includegraphics[width=\linewidth]{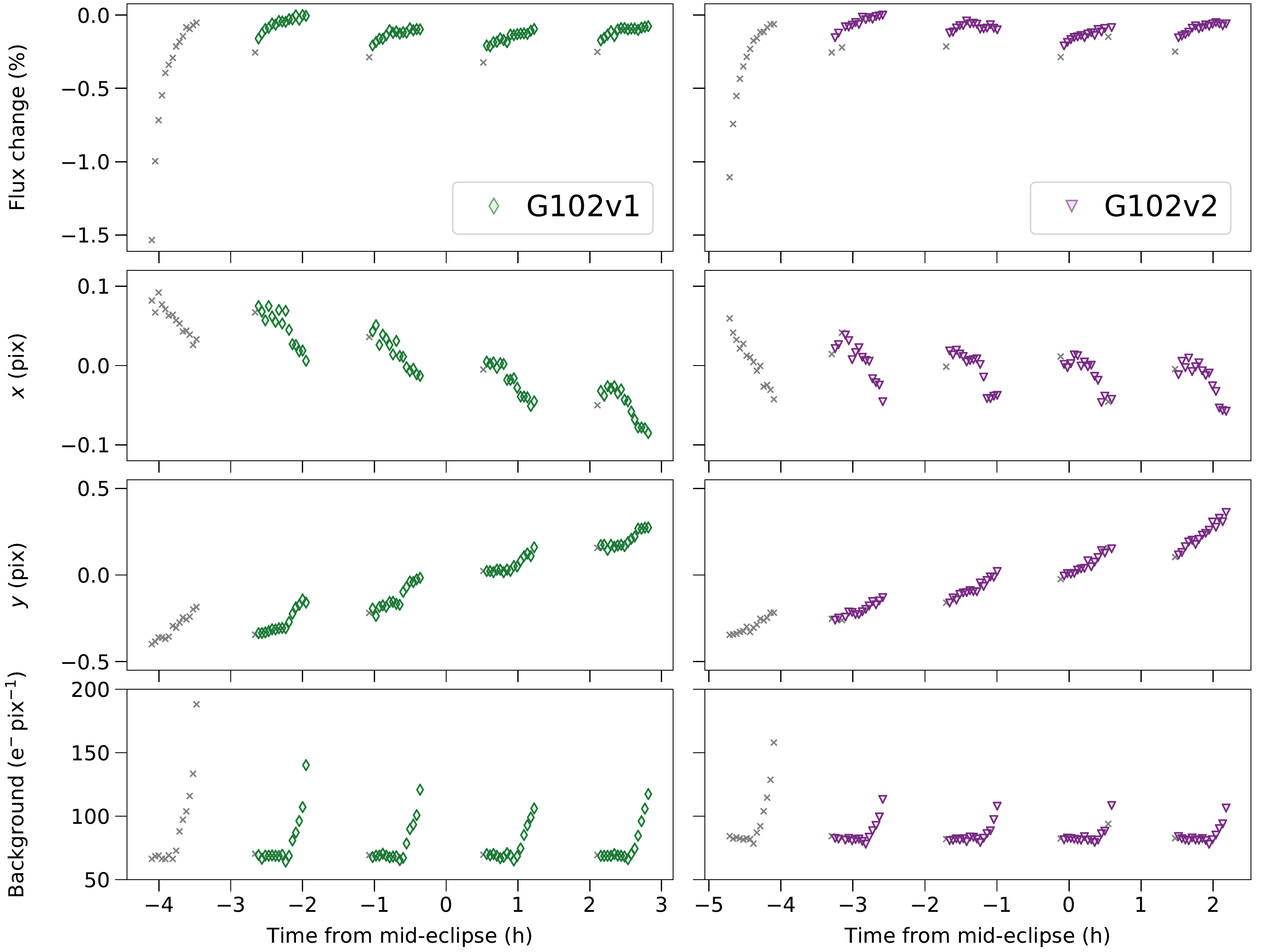}
\caption{Extracted time series for the flux, dispersion drift ($x$), cross-dispersion drift ($y$), and background level of both datasets. In all panels, colored symbols indicate data points that were included in the analysis and gray crosses indicate those that were excluded for reasons explained in the main text.}
\label{fig:timeseries}
\end{figure*}

\section{Observations and data reduction} \label{sec:observations_datared}

We observed two secondary eclipses of WASP-121b with HST/WFC3 using the G102 grism, which covers a wavelength range of approximately $0.8$-$1.1\,\um$ with a spectral resolving power of $R \sim 200$ at $\lambda = 1\,\um$ (G.O.\ 15135; P.I.\ Mikal-Evans). The first visit was made on 2017 November 6 and the second visit was made on 2017 December 9. We refer to these two visits as the G102v1 and G102v2 datasets, respectively. For both visits, the target was observed for 6.9 hours over five consecutive HST orbits with identical observing setups. The relative timing of the two visits was designed to provide full phase coverage of the eclipse, using the previously determined ephemerides of WASP-121b. Observations were made in spectroscopic mode with a forward scanning rate of $0.062$\,arcsec\,s$^{-1}$ along the cross-dispersion axis. To reduce overheads, only the $512 \times 512$ pixel subarray of the detector containing the target spectrum was read out for each exposure. We adopted the SPARS10 sampling sequence with 15 non-destructive reads per exposure ($\textnormal{NSAMP}=15$) resulting in total integration times of 103\,s and scans across approximately 50 pixel rows of the cross-dispersion axis. With this setup, we obtained 14 exposures in the first HST orbit following acquisition and 16 exposures in each subsequent HST orbit. Typical peak frame counts were $\sim 32,000$ electrons  per pixel for both visits. This translates to $\sim 13,000$\,data numbers (DN) per pixel given the detector gain of 2.5 electrons per DN, which is well within the linear regime of the WFC3 detector \citep[see Figure 1 of][]{2008wfc..rept...39H}.

Spectra were extracted from the raw data frames using a custom-built Python pipeline, which has been described previously \citep{2016ApJ...822L...4E,2017Natur.548...58E}. We started with the IMA files produced by the \textit{calwf3} pipeline version 3.4.1, which already have basic calibrations such as flat fielding, bias subtraction, and nonlinearity correction applied. The target flux was extracted from each exposure by taking the difference between successive non-destructive reads. To do this, we first estimated and subtracted the background flux for each read, by taking the median pixel count within a $10 \times 170$ pixel box which was chosen to be as large as possible while avoiding sources within the field and the detector edges. Typical background levels integrated over the full $103\,$s exposures were approximately $70$--$80$\,electrons\,pixel$^{-1}$, rising to over $100$\,electrons\,pixel$^{-1}$ at the end of each HST orbit (Figure \ref{fig:timeseries}). For each read-difference frame, we then determined the flux-weighted centre of the scanned spectrum along the cross-dispersion axis. All pixel values located more than 30 pixels above and below this row were set to zero, effectively removing flux contributions from nearby contaminant stars and cosmic ray strikes outside a rectangular aperture. Final reconstructed frames were produced by adding together the read-differences produced in this manner. During this process, we also estimated how the spectrum drifted across detector over the course of the observations. For both visits, we measure a drift of $\sim 0.1$-$0.2$\,pixel along the dispersion axis and $\sim 0.6$\,pixel along the cross-dispersion axis (Figure \ref{fig:timeseries}).

\begin{figure}
\centering  % this centres figure in column
\includegraphics[width=\columnwidth]{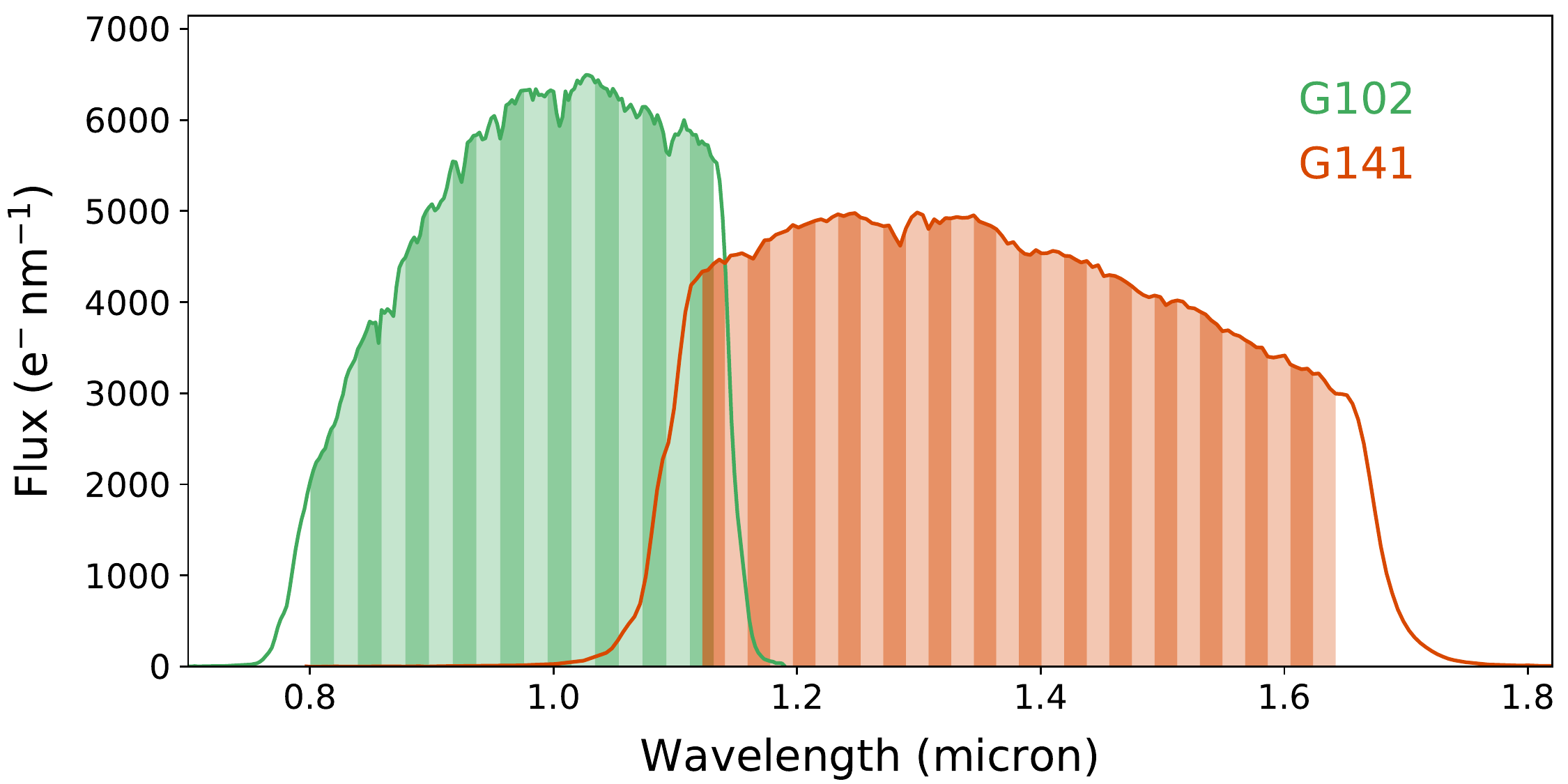}
\caption{Example spectra for the WFC3 G102 and G141 grisms. Dark and light vertical bands indicate the wavelength channels adopted for the spectroscopic light curves.}
\label{fig:example_spectra}
\end{figure}

The target spectrum was then extracted from each frame by summing the flux within a rectangular aperture spanning the full dispersion axis and 80 pixels along the cross-dispersion axis, centered on the central cross-dispersion row of the scan. The wavelength solution was determined by cross correlating each of these extracted spectra against a model spectrum for the WASP-121 host star modulated by the throughput of the G102 grism, as in \cite{2016ApJ...822L...4E,2017Natur.548...58E}. In addition to the G102 data, a single secondary eclipse of WASP-121b was observed on 2016 Nov 10 with the G141 grism (G.O.\ 14767; P.I.s Sing and L\'{o}pez-Morales). This dataset was originally published in \cite{2017Natur.548...58E}, to which the reader is referred for further details. Example G102 and G141 spectra are shown in Figure \ref{fig:example_spectra}. Together, both grisms provide continuous wavelength coverage between $\sim 0.8$-$1.6\,\um$.

\section{White Light Curve Analyses} \label{sec:whitelc}

White light curves were produced for both visits by integrating each spectrum across the full dispersion axis (Figure \ref{fig:timeseries}). The light curves are affected by the well-known hook systematic that correlates with HST orbital phase and is understood to be caused by charge trapping in the WFC3 detector \citep{2017AJ....153..243Z}. The baseline flux level also exhibits a longer term drift, which is approximately linear in time for both light curves.

Prior to light curve fitting, we chose to discard the first HST orbit of each visit, as these exhibit much stronger hooks than subsequent orbits. Although methods exist to correct the WFC3 first-orbit systematics \citep[e.g.][]{2017AJ....153..243Z,2018NatAs...2..214D}, we opted for this approach to be consistent with our previous analyses \citep{2016ApJ...822L...4E,2017Natur.548...58E} and to avoid modeling the baseline trend over the full five-orbit visits, which is less likely to be well approximated as linear. The resulting four-orbit white light curves were fit using a similar methodology to that described in \cite{2017Natur.548...58E}, in which the systematics are treated as a Gaussian process (GP). In the present study, we modeled the eclipse signal using the \texttt{batman} software package \citep{2015PASP..127.1161K}, allowing the eclipse depth ($D$) and eclipse mid-time ($\Tmid$) to vary as free parameters, while holding the remaining parameters fixed to previously determined values (Table \ref{table:whitefit}). We varied the eclipse depths jointly for both visits and allowed the mid-times to vary separately for each visit. For the GP, we employed a Mat\'{e}rn $\nu=3/2$ kernel with HST orbital phase ($\phi$), dispersion drift ($x$), and cross-dispersion drift ($y$) as input variables. We chose not to use time ($t$) as an input variable, as from past experience we have found its inclusion to make little difference to the final result \citep{2018AJ....156..283E}. The GP free parameters were the covariance amplitude ($A$) and the correlation length scales for each input variable ($L_\phi,L_x,L_y$). In practice, as in \cite{2017Natur.548...58E,2018AJ....156..283E}, we fit for the natural log of the inverse correlation length scale, $\ln\eta_k = \ln L_k^{-1}$, where $k = \{ \phi, x, y \}$. We also parameterized the white noise as $\sigma=\beta\sigma_0$, where $\sigma_0$ is the formal photon noise floor and $\beta$ is a rescaling factor that was allowed to vary in the fits. We adopted uniform priors for all eclipse parameters, and adopted the same priors as in \cite{2018AJ....156..283E} for the remaining parameters. Marginalization of the posterior distribution was performed using affine-invariant Markov chain Monte Carlo (MCMC) as implemented by the \texttt{emcee} software package \citep{2013PASP..125..306F}.

\begin{table}
\begin{minipage}{\columnwidth}
  \centering
%\scriptsize
\caption{MCMC results for the joint fit to the G102v1 and G102v2 white light curves. Quoted values are the posterior medians and uncertainties give the $\pm 34$\% credible intervals about the median. Values adopted for fixed parameters are reported at the bottom of the table. \label{table:whitefit}}
\begin{tabular}{cccc}
\hline \\ 
Free & G102v1 & G102v2 \medskip \\ \cline{1-3}
&&& \\
$D$ (ppm) & \multicolumn{2}{c}{ $682_{-73}^{+73}$ } \\
$\Tmid$ (MJD) & $58063.7624_{-0.0023}^{+0.0053}$ & $58096.9047_{-0.0015}^{+0.0024}$ \smallskip \\ 
$\beta$ & $1.21_{-0.08}^{+0.08}$ & $1.15_{-0.09}^{+0.09}$ \smallskip \\ 
$\sigma$ (ppm) & $87_{-6}^{+6}$ & $83_{-6}^{+6}$ \smallskip \\ 
\hline \\ 
Fixed & Value & Reference \medskip \\ \cline{1-3}
&&& \\
$P$ (d) & $1.2749255$ & \cite{2016MNRAS.tmp..312D} \smallskip \\ 
$\aRs$ & $3.86$ & \cite{2018AJ....156..283E} \smallskip \\ 
$i$ ($^\circ$) & $89.1$ & \cite{2018AJ....156..283E} \smallskip \\ 
$b$ & $0.06$ & \cite{2018AJ....156..283E} \smallskip \\ \hline 
\end{tabular}
\end{minipage}
\end{table}

\begin{figure}
\centering  % this centres figure in column
\includegraphics[width=0.9\columnwidth]{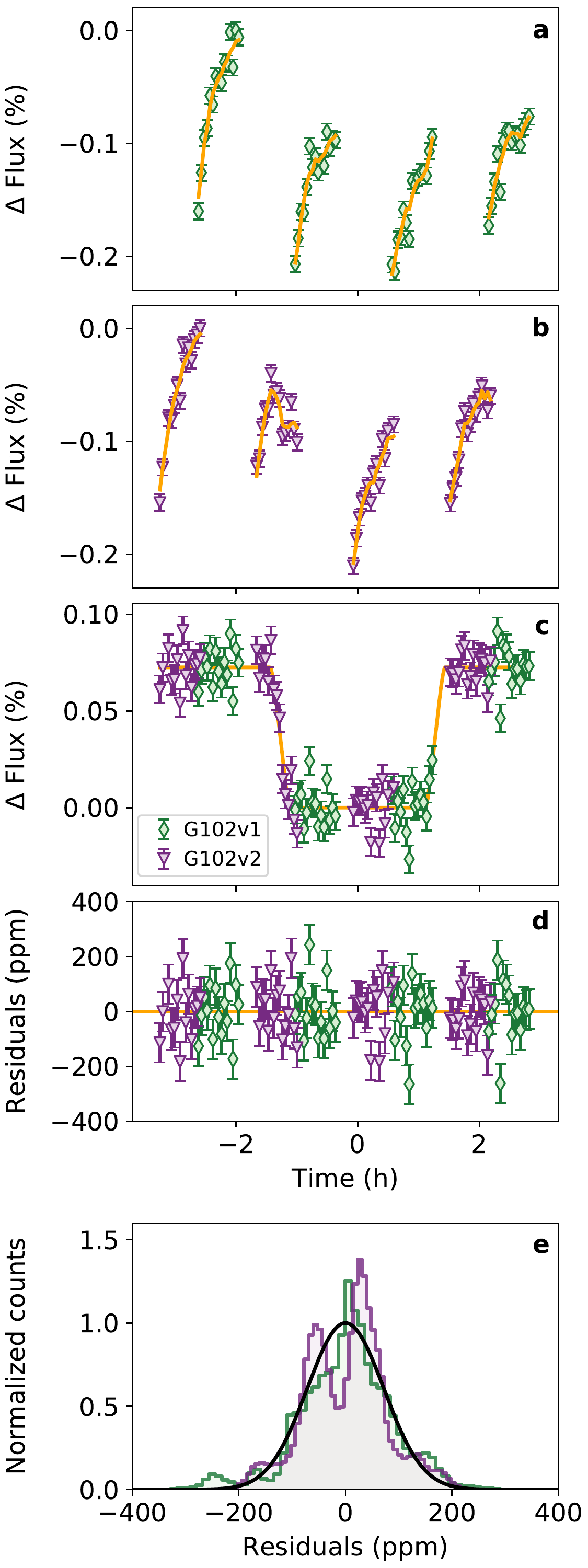}
\caption{(a) Raw white light curve for the G102v1 dataset with best-fit model indicated by orange lines and (b) the same for the G102v2 dataset. (c) Combined white light curve after removing the GP systematics component of the best-fit models, leaving only the eclipse signal. (d) Model residuals after subtracting the best-fit models from the raw light curves. (e) Normalized histograms of residuals obtained by subtracting from the data a random subset of GP mean functions obtained in the MCMC sampling. Solid black lines correspond to Gaussian distributions with standard deviations equal to photon noise (i.e.\ prior to rescaling by the $\beta$ factors described in the main text).}
\label{fig:whitefit}
\end{figure}

The resulting posterior distributions are summarized in Table \ref{table:whitefit}, and the best-fit light curve models are shown in Figure \ref{fig:whitefit}. We obtain eclipse depth measurements of $682 \pm 73$\,ppm, with inferred $\beta$ values of $1.21 \pm 0.08$ and $1.15 \pm 0.09$ for the G102v1 and G102v2 light curves, respectively. The latter imply high frequency scatter approximately $20$\,\% above the photon noise floor for both light curves, which is not accounted for by the Mat\'{e}rn kernel and is evident in the model residuals shown in Figure \ref{fig:whitefit}. As a check, we also repeated the light curve fitting using the squared exponential kernel \citep[see e.g.][]{2012MNRAS.419.2683G} and obtained results for the eclipse depth and mid-times that were fully consistent with those reported in Table \ref{table:whitefit} to well within $1\sigma$. However, the squared exponential fit gave uncertainties that were approximately 5\% smaller for the eclipse depth, 40\% smaller for the G102v1 mid-time, and 10\% smaller for the G102v2 mid-time. For this reason, we adopt the results obtained with the Mat\'{e}rn kernel to be conservative.

\section{Spectroscopic Light Curve Analyses} \label{sec:speclcs}

\begin{figure*}
\centering  % this centres figure in column
\includegraphics[width=0.7\linewidth]{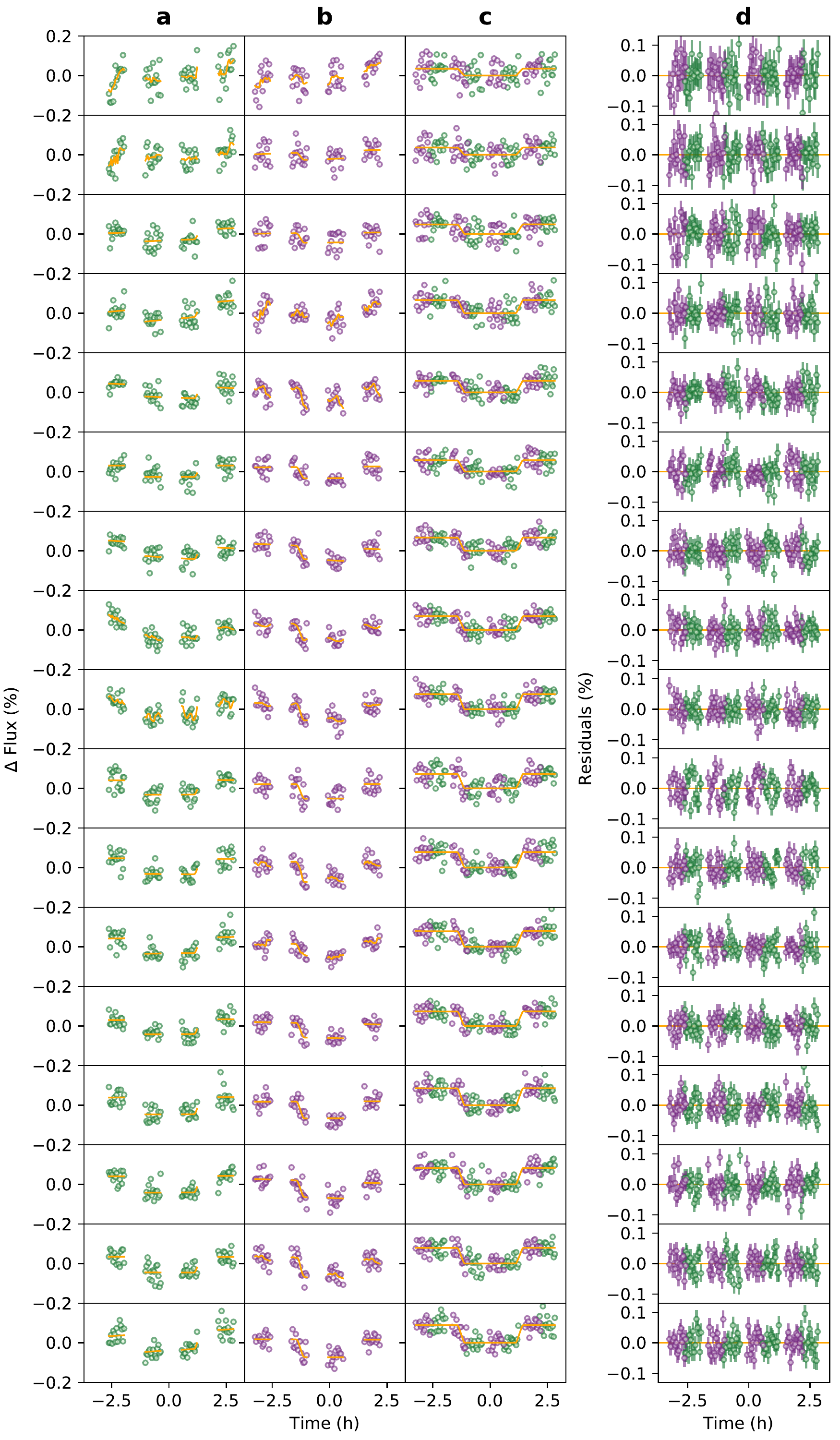}
\caption{(a) Raw spectroscopic light curves for the G102v1 dataset and (b) the same for the G102v2 dataset, with best-fit models indicated by orange lines. (c) Combined spectroscopic light curves after removing the GP systematics component of the best-fit models, leaving only the eclipse signal. (d) Residuals after subtracting the best-fit models from the raw light curves.}
\label{fig:speclcs}
\end{figure*}

Spectroscopic light curves were constructed using a similar method to that described by \cite{2013ApJ...774...95D}, which we have also used previously in \cite{2016ApJ...822L...4E,2017Natur.548...58E}. This involved cross correlating each spectrum against a master spectrum constructed by taking the median of out-of-eclipse exposures, in order to remove wavelength-independent systematics, including those arising due to pointing drift across the detector dispersion axis over the course of each visit. The flux was then binned into the 17 wavelength channels shown in Figure \ref{fig:example_spectra}, each spanning 8 pixel columns on the detector ($\Delta \lambda = 0.02 \, \um$). The resulting light curves are shown in Figure \ref{fig:speclcs}.

To fit the spectroscopic light curves, we used the same approach as described in Section \ref{sec:whitelc}. The only exception was that we fixed $\Tmid$ to the best-fit values listed in Table \ref{table:whitefit}. Thus, for the spectroscopic eclipse signals, the only free parameter was the eclipse depth $D$, which we varied jointly across both the G102v1 and G102v2 light curves. Systematics were again accounted for using GPs with Mat\'{e}rn $\nu=3/2$ kernels and white noise rescaling factors (i.e.\ $\beta$ parameters). The inferred eclipse depths and $\beta$ values are reported in Table \ref{table:specfits}.

\begin{table}
\begin{minipage}{\columnwidth}
  \centering
\scriptsize
\caption{Similar to Table \ref{table:whitefit}, but reporting the MCMC results for the joint fit to the G102v1 and G102v2 spectroscopic light curves. \label{table:specfits}}
\begin{tabular}{ccccccc}
  \hline \\
  & & \multicolumn{2}{c}{G102v1} && \multicolumn{2}{c}{G102v2}  \\ \cline{3-4} \cline{6-7}
&&&&& \\ 
  $\lambda$ ($\um$) & $D$ (ppm) & $\beta$ & $\sigma$ (ppm) && $\beta$ & $\sigma$ (ppm) \medskip \\ \cline{1-7}
&&&&& \\ 
 $0.800$-$0.820$ & $367_{-152}^{+154}$ & $1.10_{-0.08}^{+0.08}$ & $476_{-33}^{+34}$ && $1.04_{-0.08}^{+0.08}$ & $449_{-34}^{+35}$ \smallskip \\ %\smallskip \\  
$0.820$-$0.839$ & $330_{-107}^{+106}$ & $1.03_{-0.08}^{+0.08}$ & $391_{-30}^{+31}$ && $1.03_{-0.07}^{+0.07}$ & $389_{-25}^{+26}$ \smallskip \\ %\smallskip \\  
$0.839$-$0.859$ & $487_{-77}^{+81}$ & $0.99_{-0.07}^{+0.08}$ & $342_{-24}^{+27}$ && $1.09_{-0.07}^{+0.07}$ & $377_{-25}^{+26}$   \smallskip \\ %\smallskip \\  
$0.859$-$0.878$ & $657_{-81}^{+81}$ & $1.06_{-0.07}^{+0.07}$ & $350_{-23}^{+24}$ && $1.01_{-0.08}^{+0.08}$ & $330_{-25}^{+26}$   \smallskip \\ %\smallskip \\  
$0.878$-$0.898$ & $595_{-69}^{+72}$ & $1.00_{-0.07}^{+0.07}$ & $305_{-22}^{+23}$ && $0.95_{-0.08}^{+0.08}$ & $288_{-25}^{+25}$   \smallskip \\ %\smallskip \\  
$0.898$-$0.917$ & $574_{-65}^{+70}$ & $1.06_{-0.07}^{+0.08}$ & $310_{-21}^{+22}$ && $1.02_{-0.07}^{+0.07}$ & $299_{-21}^{+21}$   \smallskip \\ %\smallskip \\  
$0.917$-$0.937$ & $667_{-63}^{+67}$ & $1.06_{-0.07}^{+0.07}$ & $298_{-19}^{+21}$ && $0.99_{-0.07}^{+0.07}$ & $277_{-20}^{+21}$   \smallskip \\ %\smallskip \\  
$0.937$-$0.956$ & $671_{-83}^{+78}$ & $1.01_{-0.08}^{+0.08}$ & $277_{-21}^{+22}$ && $1.00_{-0.08}^{+0.08}$ & $274_{-21}^{+22}$   \smallskip \\ %\smallskip \\  
$0.956$-$0.976$ & $743_{-94}^{+98}$ & $1.08_{-0.10}^{+0.09}$ & $291_{-27}^{+24}$ && $1.08_{-0.08}^{+0.08}$ & $289_{-20}^{+22}$   \smallskip \\ %\smallskip \\  
$0.976$-$0.995$ & $733_{-65}^{+70}$ & $1.14_{-0.07}^{+0.07}$ & $303_{-18}^{+19}$ && $1.18_{-0.07}^{+0.07}$ & $314_{-18}^{+19}$   \smallskip \\ %\smallskip \\  
$0.995$-$1.015$ & $797_{-68}^{+64}$ & $1.12_{-0.07}^{+0.07}$ & $298_{-18}^{+19}$ && $1.03_{-0.07}^{+0.08}$ & $276_{-20}^{+21}$   \smallskip \\ %\smallskip \\  
$1.015$-$1.034$ & $795_{-67}^{+70}$ & $1.14_{-0.07}^{+0.07}$ & $298_{-18}^{+19}$ && $0.98_{-0.08}^{+0.08}$ & $258_{-20}^{+21}$   \smallskip \\ %\smallskip \\  
$1.034$-$1.054$ & $736_{-56}^{+58}$ & $1.02_{-0.07}^{+0.07}$ & $269_{-19}^{+20}$ && $1.00_{-0.07}^{+0.07}$ & $265_{-19}^{+20}$   \smallskip \\ %\smallskip \\  
$1.054$-$1.073$ & $852_{-63}^{+62}$ & $1.14_{-0.07}^{+0.07}$ & $306_{-18}^{+20}$ && $1.03_{-0.07}^{+0.08}$ & $277_{-19}^{+21}$   \smallskip \\ %\smallskip \\  
$1.073$-$1.093$ & $832_{-62}^{+62}$ & $1.04_{-0.07}^{+0.07}$ & $283_{-20}^{+20}$ && $1.10_{-0.07}^{+0.07}$ & $300_{-20}^{+20}$   \smallskip \\ %\smallskip \\  
$1.093$-$1.112$ & $791_{-65}^{+62}$ & $1.08_{-0.07}^{+0.07}$ & $299_{-19}^{+20}$ && $1.01_{-0.07}^{+0.08}$ & $279_{-20}^{+21}$   \smallskip \\ %\smallskip \\  
$1.112$-$1.132$ & $895_{-74}^{+71}$ & $1.05_{-0.08}^{+0.08}$ & $292_{-21}^{+21}$ && $1.04_{-0.07}^{+0.08}$ & $289_{-20}^{+21}$   \\ \\ \hline %\smallskip \\ \hline  
\end{tabular}
\end{minipage}
\end{table}

\begin{figure}
\centering  % this centres figure in column
\includegraphics[width=\columnwidth]{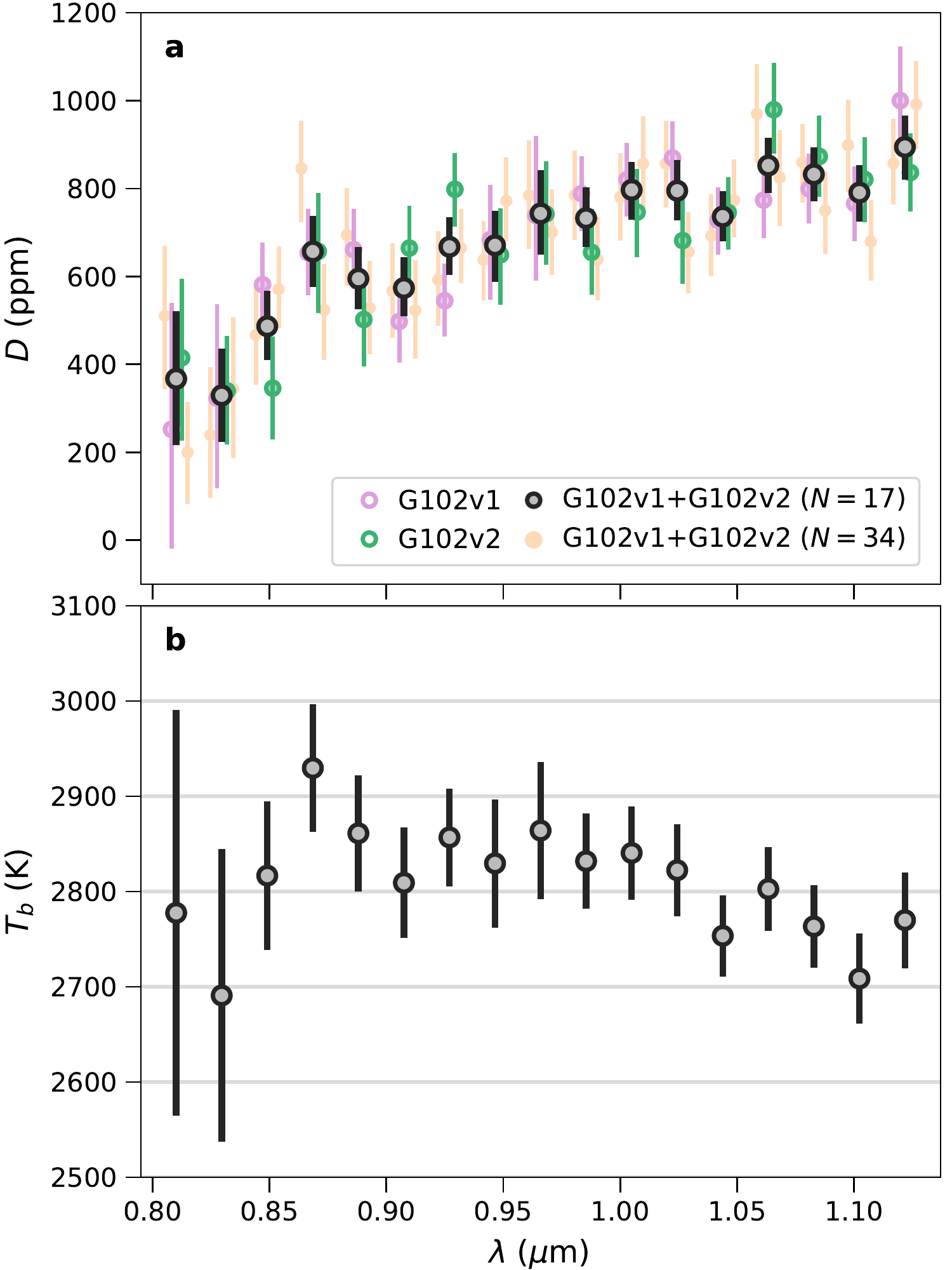}
\caption{(a) Measured eclipse depths from a joint analysis of the G102v1 and G102v2 datasets using 17 and 34 channel binnings. For the 17 channel binning, results obtained when the G102v1 and G102v2 datasets are analyzed separately are also shown. Consistent results are recovered in all cases. (b) Wavelength-dependent brightness temperatures calculated from the measured G102 eclipse depths.}
\label{fig:emspec_g102}
\end{figure}

The resulting emission spectrum is shown in Figure \ref{fig:emspec_g102}, as measured eclipse depths in the top panel and corresponding brightness temperatures in the bottom panel. Also shown are the results obtained when the G102v1 and G102v2 datasets are analyzed separately, and when the data are rebinned into 34 channels. Good agreement is obtained for all spectroscopic channels, verifying the repeatability of the measurement.

We also re-analyzed the G141 eclipse data published in \cite{2017Natur.548...58E}. To be fully consistent with the G102 analysis, we fixed the values of $\aRs$ and $i$ to the values listed in Table \ref{table:whitefit}, whereas the original G141 analysis had adopted the values reported in \cite{2016MNRAS.tmp..312D}. This gave statistically identical results to those reported in \cite{2017Natur.548...58E}, which is to be expected as $\aRs$ and $i$ primarily affect the eclipse duration, rather than the eclipse depth. 

\section{Discussion} \label{sec:discussion}

\begin{figure}
\centering  % this centres figure in column
\includegraphics[width=\columnwidth]{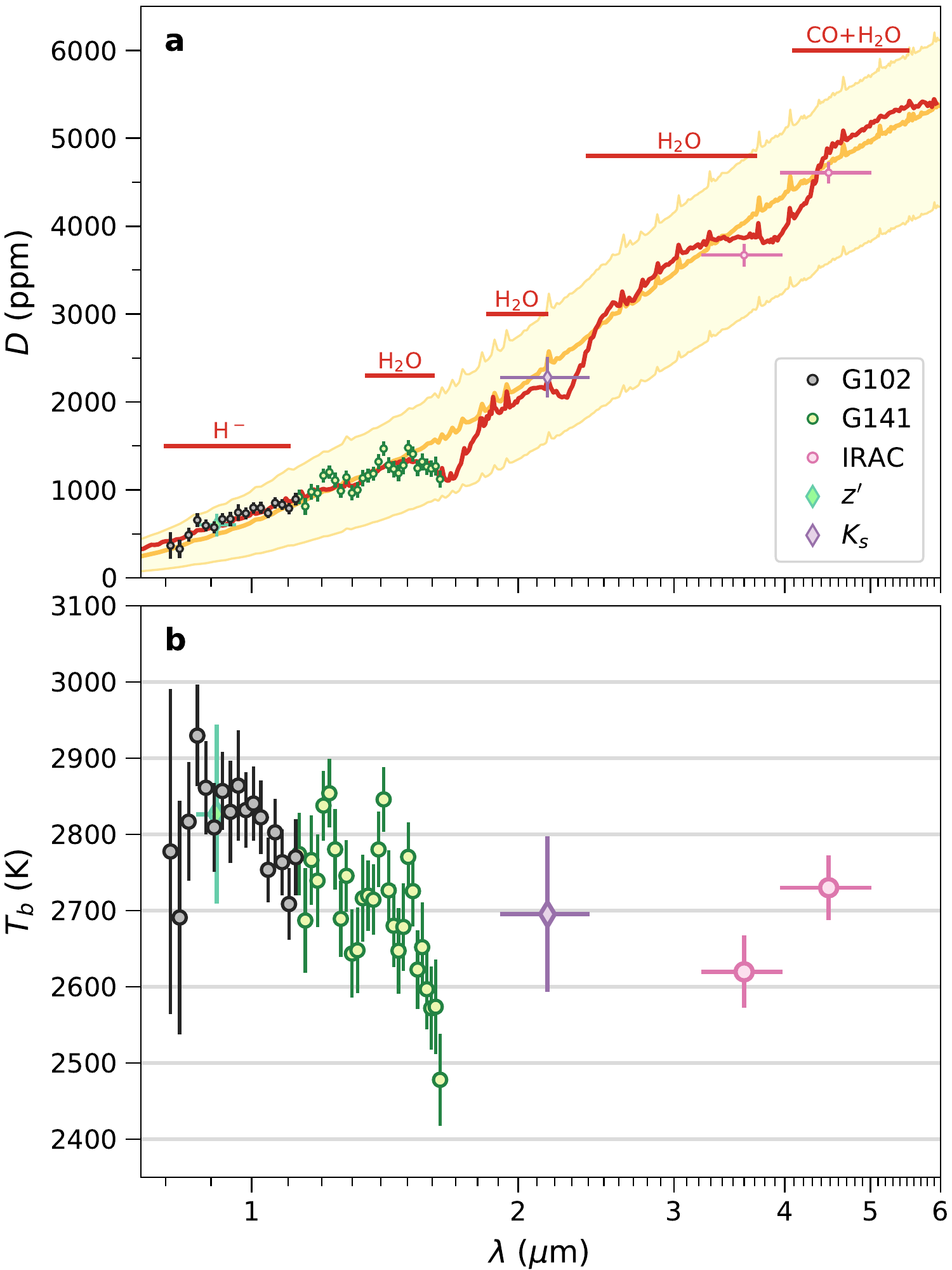}
\caption{(a) Published eclipse depth measurements for WASP-121b. Dark yellow line shows the expected spectrum if the planet were to radiate as a blackbody at the best-fit temperature of 2720\,K. Pale yellow envelope indicates a plausible range of emission assuming the planet has zero albedo and radiates as a blackbody, with the lower limit corresponding to uniform emission from the dayside and nightside hemispheres at a temperature of 2330\,K, and the upper limit corresponding to emission from the dayside only at a temperature of 2970\,K. Red line shows the best-fit model from the retrieval analysis, with spectral emission features due to H$^-$, H$_2$O, and CO labeled. (b) Brightness temperatures derived from the measured eclipse depths. These correspond to the temperatures required to match the implied fluxes in each channel if the planet were to radiate as a blackbody.}
%\caption{Eclipse depth measurements for WASP-121b. Dark yellow line shows the expected spectrum if the planet were to radiate as a blackbody at the best-fit temperature of 2720\,K. Pale yellow envelope indicates a plausible range of emission assuming the planet has zero albedo and radiates as a blackbody, with the lower limit corresponding to uniform emission from the dayside and nightside hemispheres at a temperature of 2330\,K, and the upper limit corresponding to emission from the dayside only at a temperature of 2970\,K. Red line shows the best-fit model from the retrieval analysis, with spectral emission features due to H$^-$, H$_2$O, and CO labeled.}
\label{fig:emspec_all}
\end{figure}

The secondary eclipse spectrum measured to date for WASP-121b is shown in Figure \ref{fig:emspec_all}, along with the corresponding brightness temperatures. In addition to the G102 data spanning $0.8$-$1.1\,\um$ presented in the current study, this includes the G141 data spanning $1.1$-$1.6\,\um$ from \cite{2017Natur.548...58E}, ground-based photometric measurements in the $z^\prime$ \citep{2016MNRAS.tmp..312D} and $K_s$ \citep{2019A&A...625A..80K} passbands, and the IRAC data at $3.6\,\um$ and $4.5\,\um$ from \cite{2019arXiv190107040G}. The G102 and G141 measurements agree extremely well at the point of overlap between the two passbands, without any adjustment to the level of either dataset. Similarly, the G102 data are fully consistent with the $z^\prime$ measurement without any adjustment.

\subsection{Blackbody fits and heat redistribution} \label{sec:discussion:blackbody}

% see figures/figures.py EmSpecFigs() routine for printouts of blackbody fits.

To interpret the data, we first consider the simple case in which the planet is assumed to radiate as an isothermal blackbody. Fitting such a model to the full dataset gives a best-fit temperature of $2720 \pm 8$\,K, with predictions for the wavelength-dependent secondary eclipse depth indicated by the dark yellow line in Figure \ref{fig:emspec_all}. The data approximately follow the shape of this curve, however, the reduced $\chi^2$ is 2.92 for 47 degrees of freedom, allowing it to be ruled out at $6.5\sigma$ confidence. For comparison, in \cite{2017Natur.548...58E} we ruled out a blackbody model at $5\,\sigma$ confidence by fitting to the data available at that time; namely, the G141, $z^\prime$, and IRAC $3.6\,\um$ measurements. The addition of the G102 and IRAC $4.5\,\um$ measurements has therefore increased the discrepancy between the data and a blackbody model.

We also experimented with fitting a blackbody model to different subsets of the data. For instance, if we repeat the fit to the full dataset with the only exception being that we use the G141 white eclipse depth rather than the G141 spectroscopic eclipse depths, we obtain a best-fit temperature of $2754 \pm 11$\,K with a reduced $\chi^2$ of 1.97 for 21 degrees of freedom, reducing the confidence with which such a model can be ruled out to $2.8\sigma$. Alternatively, if we repeat the fit to the full dataset but exclude the G102 spectroscopic eclipse depths --- similar to \cite{2017Natur.548...58E} but with the addition of the $K_s$ and IRAC $4.5\,\um$ points --- we obtain a best-fit temperature of $2691 \pm 9$\,K with a reduced $\chi^2$ of 2.88 for 30 degrees of freedom, ruling it out at $5.2\sigma$ confidence. 

These results imply that, statistically, the departure from a blackbody spectrum is largely driven by the spectroscopic information contained in the G141 data. In \cite{2017Natur.548...58E}, we attributed this departure to a muted H$_2$O emission band at $1.4\,\um$, as well as a tentative VO emission band at $1.25\,\um$. We revisit this interpretation in Section \ref{sec:discussion:modeling} with a retrieval analysis of the updated dataset. Here we note that even without the G141 spectroscopic information, considerable tension remains between the data and a blackbody model, mainly due to the mismatch between the overall slope of the data and a blackbody spectrum. This can be appreciated in Figure \ref{fig:emspec_all}, which shows a systematic decrease in brightness temperature over the near-infrared wavelength range covered by the G102 and G141 passbands. As we discuss in Section \ref{sec:discussion:modeling}, much of this variation can be explained by the $1.4\,\um$ H$_2$O band, as well as continuum opacity due to H$^-$ ions.

\begin{figure}
\centering  % this centres figure in column
\includegraphics[width=\columnwidth]{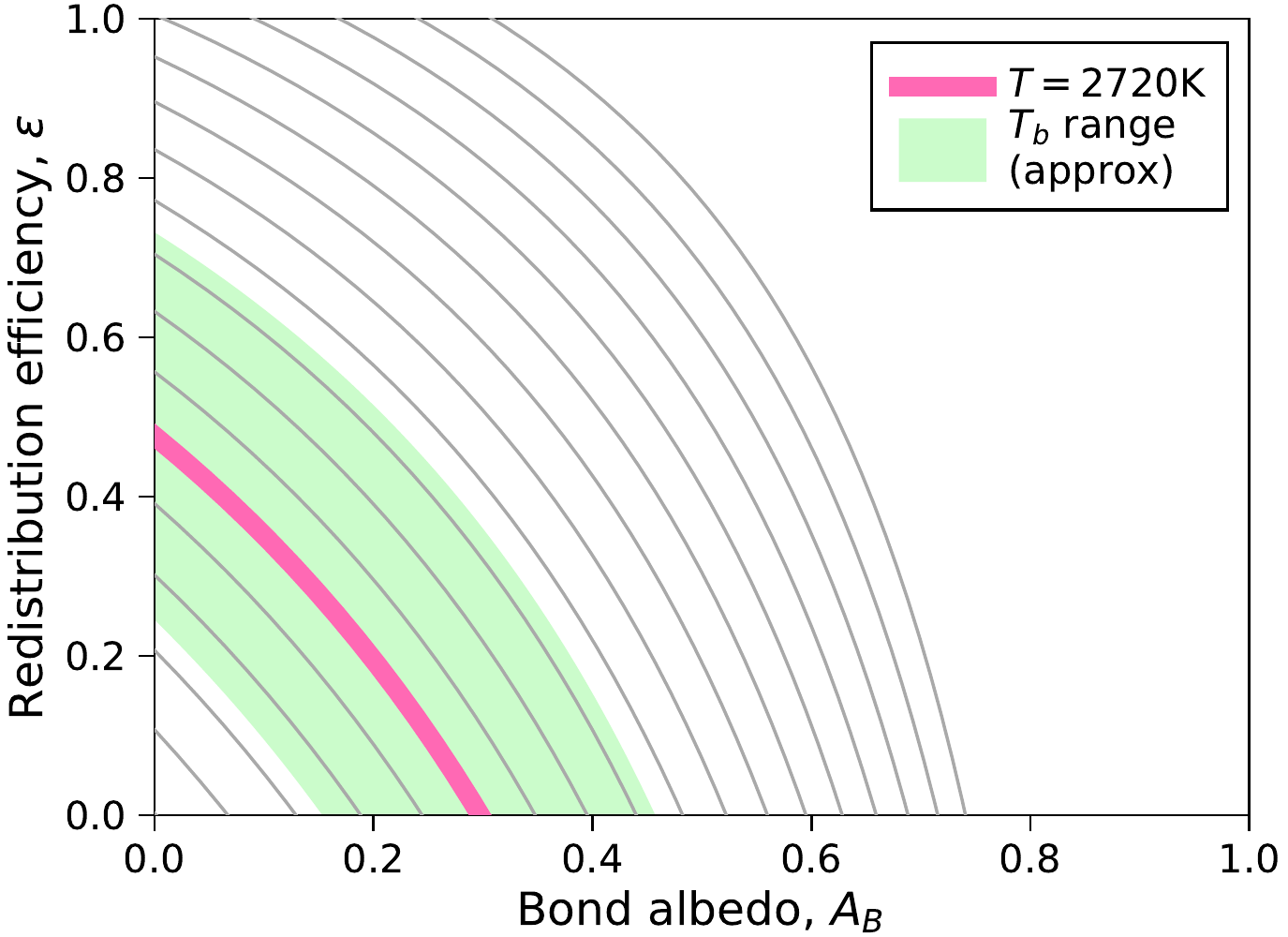}
\caption{Degeneracy between heat redistribution efficiency and Bond albedo given the temperature of the dayside hemisphere. Thick pink line shows allowed values assuming the dayside radiates as a blackbody with best-fit temperature 2720\,K. Pale green region indicates the allowed values implied by the range of brightness temperatures shown in Figure \ref{fig:emspec_all}. Thin gray lines show contours for different dayside temperatures separated by 50\,K increments. }
\label{fig:recircalbedo}
\end{figure}

Figure \ref{fig:recircalbedo} shows the approximate range of Bond albedos ($A_B$) and heat redistribution efficiencies ($\varepsilon$) for WASP-121b allowed by the emission data. The $\varepsilon$ parameter is defined following \cite{2011ApJ...729...54C}, with $\varepsilon=0$ corresponding to zero heat redistribution and $\varepsilon=1$ corresponding to uniform heat redistribution. Measurements of low Bond albedos for numerous hot Jupiters \citep[e.g.][]{2011MNRAS.417L..88K,2013ApJ...777..100H,2017ApJ...847L...2B,2018AJ....156...44M,2019A&A...624A..62M} suggest a low Bond albedo is also likely for WASP-121b, which would be in line with the evidence for significant optical absorption in the transmission spectrum \citep{2018AJ....156..283E}. Under this scenario ($A_B\lesssim 0.1$), the heat redistribution efficiency would be $\varepsilon \approx 0.4$. However, if heat redistribution is inefficient ($\varepsilon \approx 0$), the maximum allowable Bond albedo is $A_B \approx 0.3$. A phase curve measurement would allow this degeneracy to be broken, by providing a direct constraint on the heat flux from the nightside hemisphere, and hence on $\varepsilon$.

%\subsection{Retrieval analysis} \label{sec:discussion:retrieval}
%\subsection{Forward modeling} \label{sec:discussion:forward}
\subsection{Atmosphere modeling of the dayside hemisphere} \label{sec:discussion:modeling}

\begin{figure}
\centering  % this centres figure in column
\includegraphics[width=\columnwidth]{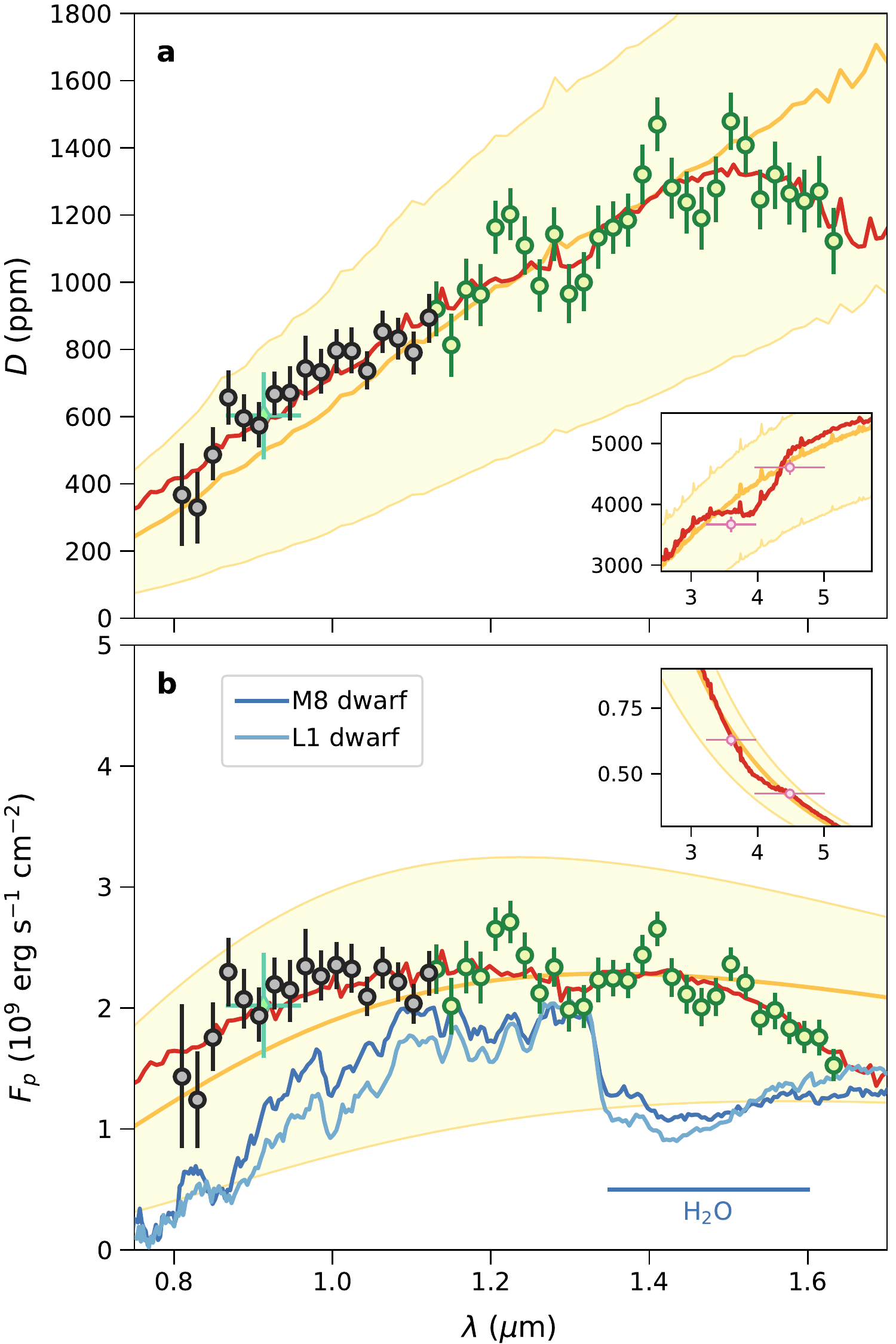}
\caption{(a) Similar to panel (a) of Figure \ref{fig:emspec_all}, but with a linear horizontal scale and inset showing the IRAC data to allow closer inspection of the measured eclipse depths and models. (b) Planet surface flux derived from the eclipse depths shown in panel (a). Blue lines show measured M8 and L1 dwarf spectra \citep{2010ApJS..190..100K}, with arbitrary normalization applied. These objects have comparable photospheric temperatures to WASP-121b and exhibit deep H$_2$O absorption bands at $1.4\,\um$, as they do not have thermal inversions at the photosphere.}
\label{fig:emspec_all_lin}
\end{figure}

We performed an atmospheric retrieval analysis for the secondary eclipse data shown in Figure \ref{fig:emspec_all_lin} using the \texttt{ATMO} code, which has been described extensively elsewhere \citep{2014A&A...564A..59A,2015ApJ...804L..17T,2016ApJ...817L..19T,2017ApJ...841...30T,2016A&A...594A..69D,2018MNRAS.474.5158G}. In brief, \texttt{ATMO} was originally developed to solve the one-dimensional (1D) plane-parallel radiative transfer equation assuming hydrostatic balance and radiative-convective equilibrium, with a two-dimensional implementation subsequently introduced by \cite{2017ApJ...841...30T}. For a given atmospheric composition and pressure-temperature (PT) profile, chemical equilibrium gas phase abundances can be determined by Gibbs energy minimization, or alternatively, arbitrary mixing ratios can be specified. The PT profile can also be provided as an input, or otherwise computed self-consistently given the atmospheric opacity sources, an internal heat flux, and the irradiation from the host star. Condensation can be treated either locally or using a rainout approach \citep[see][]{2019MNRAS.482.4503G}. The planetary spectrum viewed by an external observer is produced as output, allowing \texttt{ATMO} to be used for inferring atmospheric properties from primary transit \citep{2016ApJ...822L...4E,2018AJ....156..283E,2017Sci...356..628W,2018AJ....155...29W,2018Natur.557..526N,2018AJ....156..298A} and secondary eclipse \citep{2017Natur.548...58E,2018MNRAS.474.1705N} measurements.

\begin{figure*}
\centering  % this centres figure in column
\includegraphics[width=\linewidth]{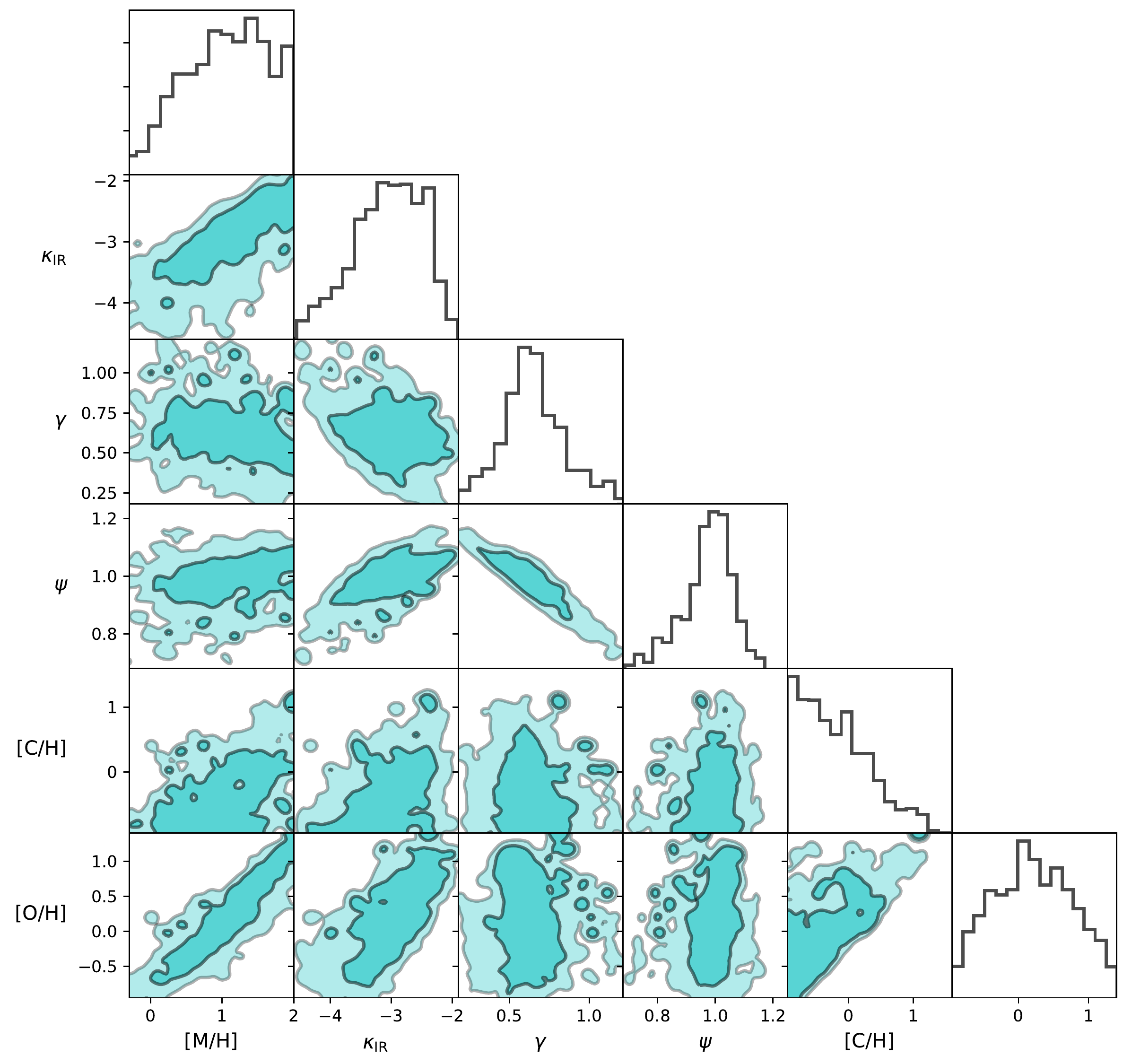}
\caption{Posterior distributions for the free parameters of the retrieval analysis. Panels along the diagonal show the marginalized distributions for each individual parameter. Off-diagonal panels show the marginalized distributions for parameter pairs, with contours indicating the 68\% and 95\% credible ranges. }
\label{fig:posterior}
\end{figure*}

We previously used \texttt{ATMO} in \cite{2017Natur.548...58E} to analyze the G141, $z^\prime$, and IRAC $3.6\,\um$ emission data for WASP-121b. In that study, we assumed uniform mixing ratios with pressure and allowed the abundances of H$_2$O and VO to vary as free parameters. In the present study, we add the G102, $K_s$, and IRAC $4.5\,\um$ data to our analysis and make a number of changes to our retrieval methodology, motivated by recent work highlighting the importance of thermal dissociation and ionization in ultrahot Jupiter atmospheres \citep{2018ApJ...855L..30A,2018A&A...617A.110P,2018ApJ...866...27L,2018AJ....156...17K,2018AJ....156...10M}. First, rather than fitting for unconstrained abundances of pre-defined opacity sources, we assumed chemical equilibrium with relative elemental abundances set to solar values and varied metallicity ($\Z$), as well as the carbon and oxygen elemental abundances ($\Cabun$, $\Oabun$) separately.\footnote{Under this formulation, $\Z$ controls the number density of all elements heavier than helium except for carbon and oxygen.} Second, we accounted for condensation using the rainout scheme described in \cite{2019MNRAS.482.4503G} and included the effect of gas phase scattering. Third, we allowed for thermal dissociation and ionization of atoms and molecules, which can result in strongly pressure-dependent abundances for many chemical species throughout layers of the atmosphere probed in emission. We note that \texttt{ATMO} has always accounted for these effects in determining chemical equilibrium abundances of neutral species, but this was not included in our previous retrieval, as we assumed a uniform distribution of H$_2$O and VO in pressure and allowed the abundances to vary freely. Ionic species were, however, added to \texttt{ATMO} for the present study following \cite{1994GordonMcBride}, and the ion-neutral composition was benchmarked against the open source GGChem code \citep{2018A&A...614A...1W}. The resulting chemical system consisted of 175 neutral gas phase species, 93 condensates, and the ionized species e$^-$, H$^+$, H$^-$, He$^+$, Na$^+$, K$^+$, C$^+$, Ca$^+$, and Si$^+$. The most important radiatively active neutral gas phase species were H$_2$O, CO$_2$, CO, CH$_4$, NH$_3$, Na, K, Li, R, Cs, TiO, VO, FeH, PH$_3$, H$_2$S, HCN, C$_2$H$_2$, SO$_2$, and Fe(g). Collision-induced absorption due to H$_2$-H$_2$ and H$_2$-He was also included.
  
As in \cite{2017Natur.548...58E}, we fit for the PT profile using the analytic solution derived by \cite{2010A&A...520A..27G}, which is parameterized in terms of the infrared opacity ($\kir$), the ratio of the visible-to-infrared opacity ($\gamma=\kappa_{\textnormal{V}}/\kappa_{\textnormal{IR}}$), and an irradiation efficiency factor ($\psi$). The latter is identical to the $\beta$ term defined by \cite{2013ApJ...775..137L}, but we denote it as $\psi$ to avoid confusion with the white noise rescaling factors used in Sections \ref{sec:whitelc} and \ref{sec:speclcs}. It effectively allows for nonzero albedo values and varying degrees of heat recirculation from the dayside to nightside hemisphere.

\begin{table}
\begin{minipage}{\columnwidth}
  \centering
%\scriptsize
\caption{Retrieval prior ranges and MCMC results \label{table:retrieval}}
\begin{tabular}{cccc}
 \hline \\ 
 Parameter & Unit & Allowed range & Result  \medskip \\ \cline{1-4}
 &&& \\
 $\Z$ & dex &                              $-1$ to $2$ & ${1.09}_{-0.69}^{+0.57}$ \smallskip \\
 $\Cabun$ & dex &                          $-1$ to $2$ & ${-0.29}_{-0.48}^{+0.61}$ \smallskip \\
 $\Oabun$ & dex &                          $-1$ to $2$ & ${0.18}_{-0.60}^{+0.64}$ \smallskip \\
 $\log_{10}(\kir)$ & dex\,cm$^2$\,g$^{-1}$ & $-5$ to $0.5$ & ${-3.01}_{-0.62}^{+0.56}$ \smallskip \\
 $\log_{10}(\gamma)$ & dex &                $-4$ to $1.5$ & ${0.64}_{-0.16}^{+0.19}$ \smallskip \\
 $\psi$ & ---  &                            $0$ to $2$   & ${0.99}_{-0.09}^{+0.06}$ \\ \vspace{-4pt} \\ \hline  

 \end{tabular}
\end{minipage}
\end{table}

We adopted uniform priors for all six retrieval parameters (i.e. $\Z$, $\Cabun$, $\Oabun$, $\log_{10}\kir$, $\log_{10}\gamma$, $\psi$), with the allowed ranges listed in Table \ref{table:retrieval}. Differential-evolution MCMC was used to marginalize the posterior distribution, using the publicly available software of \cite{2013PASP..125...83E}. We ran twelve chains each for 30,000 steps and discarded the first 20\% of each chain as burn-in before combining them into a single chain. The resulting parameter distributions are reported in Table \ref{table:retrieval} and shown in Figure \ref{fig:posterior}. The best-fit emission spectrum is shown in Figure \ref{fig:emspec_all_lin} and has a $\chi^2$ of 43.61 for $\nu=42$ degrees of freedom, indicating an excellent fit to the data with reduced $\chi^2_\nu=1.04$. %The inferred pressure-dependent abundances, PT profile, and contribution functions are shown in Figure \ref{fig:vertresults}.

\begin{figure}
\centering  % this centres figure in column
\includegraphics[width=\columnwidth]{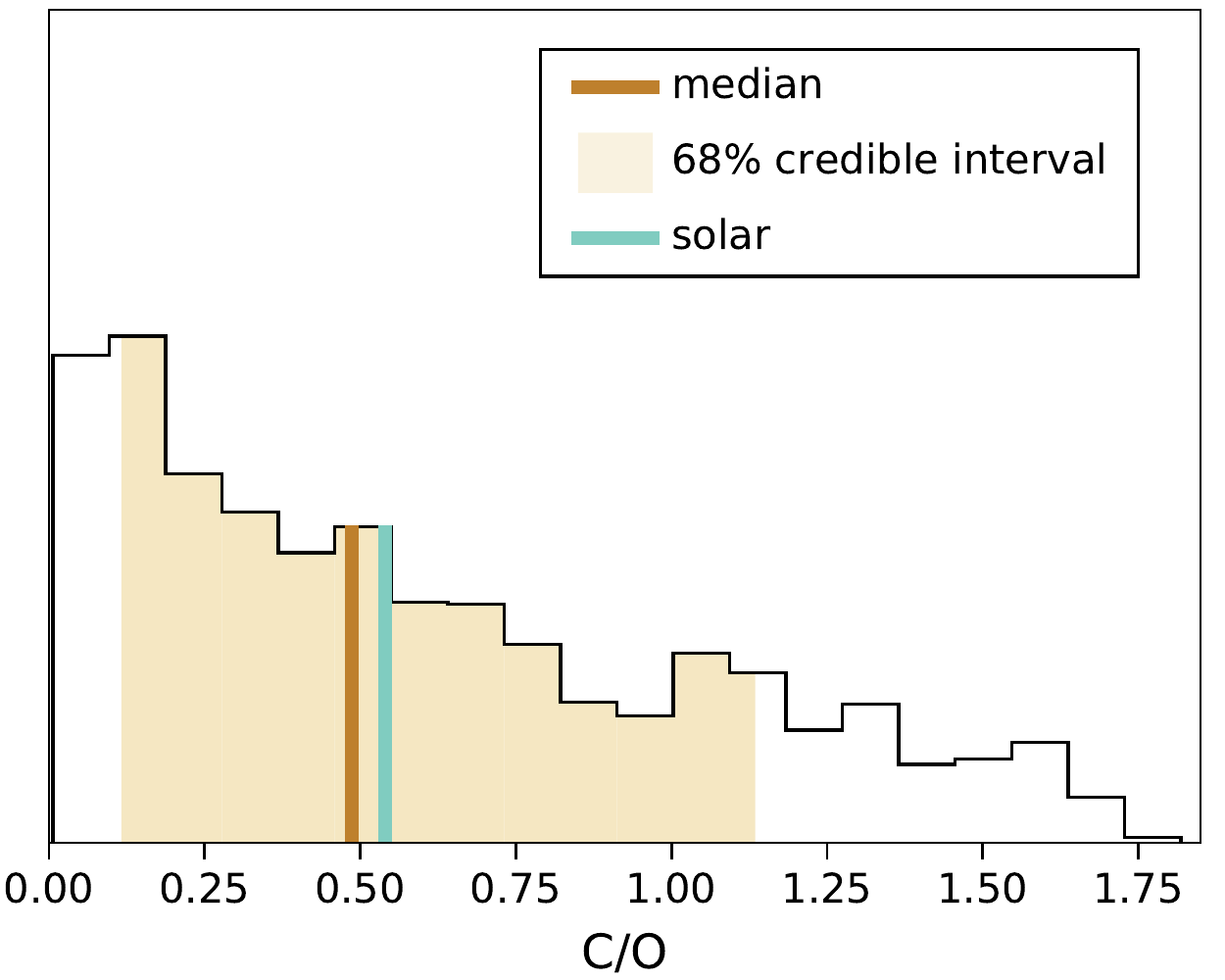}
\caption{Normalized posterior distribution for the carbon-to-oxygen ratio, obtained by combining the $\Cabun$ and $\Oabun$ samples shown in Figure \ref{fig:posterior}.}
\label{fig:co}
\end{figure}

\begin{figure*}
\centering  % this centres figure in column
\includegraphics[width=\linewidth]{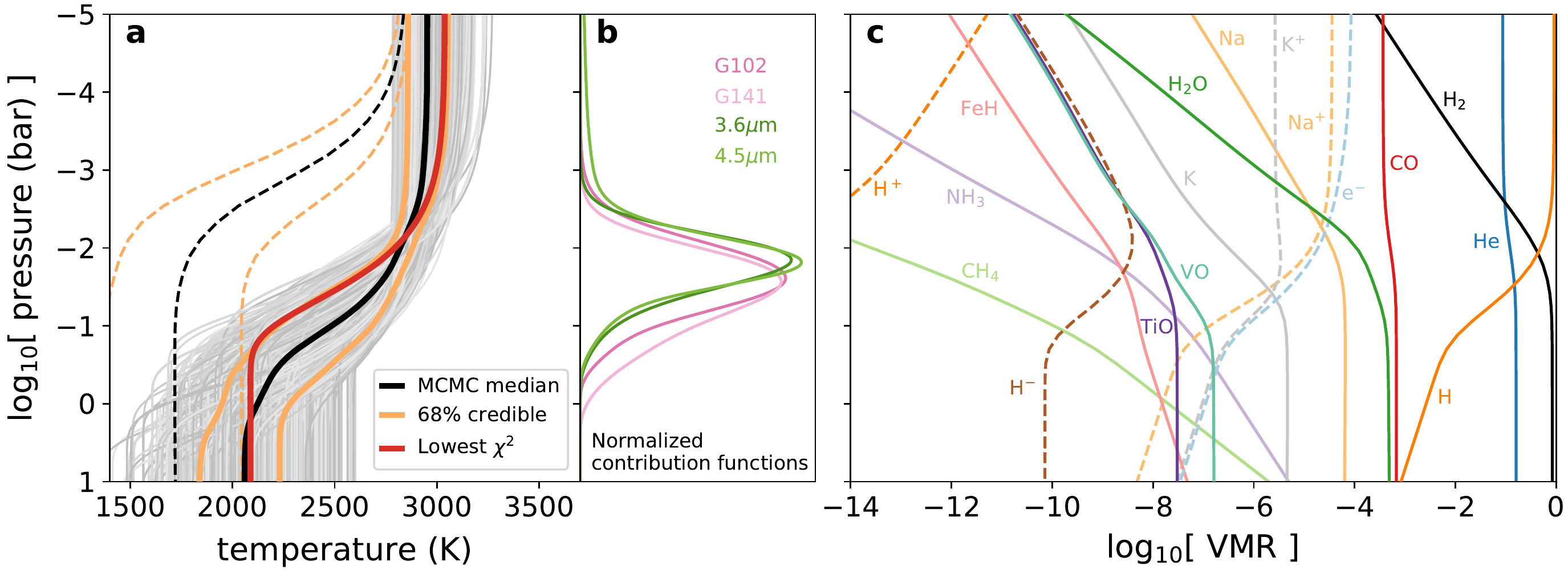}
\caption{(a) PT profile inferred from the retrieval analysis, which uses the analytic solution referred to in the main text. Gray lines show a subset of the MCMC samples, black line shows the median temperature at each pressure level, and dark yellow lines demarcate the 68\% credible interval. Red line shows the PT profile for the best-fit model plotted in Figures \ref{fig:emspec_all} and \ref{fig:emspec_all_lin}. Dashed lines show the PT distribution presented previously in \protect\cite{2017Natur.548...58E} - see text for discussion. (b) Contribution functions for the best-fit model, integrated over the different HST and \textit{Spitzer} passbands. (c) Pressure-dependent abundances for the best-fit model. Important opacity sources such as H$_2$O, TiO, VO, and FeH are heavily depleted for pressures less than $\sim 50$\,mbar due to thermal dissociation, whereas CO is relatively unaffected. Note also that VO is more abundant than TiO at pressures greater than $\sim 10\,$mbar, because for the specific model shown here TiO has condensed and partially rained out but VO has not. For other PT profiles sampled from the posterior, TiO did not condense and had a higher abundance than VO, in line with the relative Ti and V abundances of the Sun. Given that the available data does not show robust evidence for TiO or VO spectral features, these differences did not affect the fit quality.}
\label{fig:vertresults}
\end{figure*}

We infer a metallicity of $\Z = {1.09}_{-0.69}^{+0.57}$, translating to a 68\% credible interval of $\sim 5$-$50\times$ solar. This is consistent with the plausible metallicity range of $\sim 10$-$30\times$ solar we found for the transmission spectrum \citep{2018AJ....156..283E}. Note that the WASP-121 host star itself shows only mild evidence for heavy metal enrichment relative to solar, with $[\textnormal{Fe}/\textnormal{H}]=0.13 \pm 0.09$ \citep{2016MNRAS.tmp..312D}. We also obtain 68\% credible intervals of $-0.77$ to $0.33$ for $\Cabun$ and $-0.42$ to $0.82$ for $\Oabun$. These latter ranges are somewhat lower than that inferred for $\Z$, but are within an order of magnitude. Figure \ref{fig:co} shows the corresponding carbon-to-oxygen ratio, obtained by combining the $\Cabun$ and $\Oabun$ MCMC chains. We estimate $\textnormal{C/O} = 0.49_{-0.37}^{+0.65}$, which is consistent with the solar value of $0.54$ \citep{2009ARA&A..47..481A}. Together, the updated emission spectrum presented here and the transmission spectrum presented in \cite{2018AJ....156..283E} suggest a metallicity of $\sim 5$-$50\times$ solar for the atmosphere of WASP-121b at pressures below $\sim 100$\,mbar, while the C/O ratio is weakly constrained but fully consistent with that measured for the Sun.

These results are broadly in line with theoretical predictions for the heavy element content of gas giant atmospheres. For example, the interior structure models of \cite{2019ApJ...874L..31T} predict a metallicity of $21 \pm 4 \ \times$ solar for the upper atmosphere of WASP-121b, well within the ranges favored by the transmission and emission spectra.  However, our revised metallicity estimate is in contrast to the anomalously high abundances for H$_2$O and VO reported in \cite{2017Natur.548...58E} for the dayside atmosphere of WASP-121b. In that study we obtained a 95\% credible lower limit of $1,000\times$ solar for the VO abundance, which was driven by the apparent flux excess measured across the $1.20$-$1.25\,\um$ wavelength range. Although the equilibrium chemistry model presented here fails to replicate this feature (Figure \ref{fig:emspec_all_lin}), the overall goodness of fit it provides to the full dataset (i.e. $\chi^2_\nu=1.04$) suggests it is more likely a statistical fluctuation than a VO emission band.

The ability of the chemical equilibrium model to fit the data with abundances closer to solar values is primarily due to the effects of thermal dissociation and ionization, and the opacities of the resulting products such as H$^-$\textbf, which as noted above were not treated in our previous retrieval analysis. \cite{2018A&A...617A.110P} were the first to include these effects in a detailed analysis of the WASP-121b dayside emission data and made a similar observation. Those authors employed a three-dimensional (3D) general circulation model (GCM), which has the advantage of treating the radiative transfer and dynamics of the atmosphere self-consistently. However, a disadvantage of 3D GCMs is their computational expense, which typically prevents the free parameters (e.g.\ metallicity, frictional drag, etc.) being optimized to match a given dataset. This practical limitation may explain the somewhat poorer match to the data provided by the GCM of Parmentier et al.\ compared to that obtained by our retrieval analysis shown in Figure \ref{fig:emspec_all_lin}.

While we stress that GCMs represent the state-of-the-art for modelling the 3D interplay between circulation, radiation, and chemistry, by implementing a simpler model, our retrieval analysis is able to more fully explore its parameter space and optimize the match to the data. This comes at the costs of approximating the atmosphere as 1D, not solving for the PT profile self-consistently given the atmospheric opacity, and ignoring dynamical effects. For each MCMC sample, the retrieval instead simply takes a step in the six-dimensional parameter space and uses the resulting values for $\kir$, $\gamma$, and $\psi$ to evaluate the PT profile, and the values of $\Z$, $\Cabun$, and $\Oabun$ to determine the elemental abundances. These two inputs --- the PT profile and elemental abundances --- are then used to solve for the chemical equilibrium abundances, with the effects of thermal dissociation and ionization accounted for appropriately.

Figure \ref{fig:vertresults} shows the PT profile, contribution functions, and abundances obtained in this way from our retrieval analysis. Specifically, the results shown are for the best-fit model (i.e.\ the MCMC sample with the lowest $\chi^2$ value), with panel (a) also displaying the distribution of PT profiles across all MCMC samples. As in \cite{2017Natur.548...58E}, we find the PT profile exhibits an unambiguous thermal inversion. However, our new retrieval analysis puts the $\sim 2700$\,K photosphere much deeper within the atmosphere than implied by the PT distribution presented in \cite{2017Natur.548...58E}, which is indicated by dashed lines in panel (a) of Figure \ref{fig:vertresults}. For the latter, $2700$\,K coincides with pressures $\sim 10\,\mu$bar, versus $\sim 10$\,mbar for the updated PT profile. This is a consequence of the different modeling assumptions made in the two studies, as detailed above. In particular, the \cite{2017Natur.548...58E} analysis did not enforce chemical equilibrium, and consequently dialled the VO abundance up to unrealistically high values in order to fit the $1.25\um$ bump in the measured spectrum. Since the VO abundance was assumed to be uniform across all pressure levels, this in turn increased the opacity throughout the entire atmosphere column, explaining why the photosphere occurs at such low pressures for that model. In the updated retrieval analysis, the abundances of important absorbers such as H$_2$O, VO, and TiO are greatly reduced at low pressures due to thermal dissociation (Figure \ref{fig:vertresults}), moving the photosphere to higher pressures.

We find that the temperature increases from approximately $2500$\,K to $2800$\,K across the pressures probed by the data (i.e.\ $\sim 30$ to $5$\,mbar for the best-fit model as seen in Figure \ref{fig:vertresults}), which is close to the range spanned by the brightness temperatures shown in Figure \ref{fig:emspec_all}. The inference of a thermal inversion is driven by various spectral features in the data appearing in emission rather than absorption, notably: H$^-$ within the G102 passband; H$_2$O within the G141 passband; and unresolved CO and H$_2$O bands within the IRAC $4.5\,\um$ channel (Figures \ref{fig:emspec_all} and \ref{fig:emspec_all_lin}). As others have highlighted \citep[][]{2018ApJ...855L..30A,2018ApJ...866...27L,2018A&A...617A.110P,2018AJ....156...17K,2018AJ....156...10M}, the H$_2$O bands are significantly muted due to thermal dissociation in the upper layers of the atmosphere, owing to the intense irradiation by the nearby host star. This can be appreciated by examining panel (c) of Figure \ref{fig:vertresults}, which shows the effect of thermal dissociation and ionization for various species in our best-fit model. For H$_2$O, the volume mixing ratio decreases by a factor of $\sim 300$ between pressures of 100 mbar and 1 mbar.

\begin{figure}
\centering  % this centres figure in column
\includegraphics[width=\columnwidth]{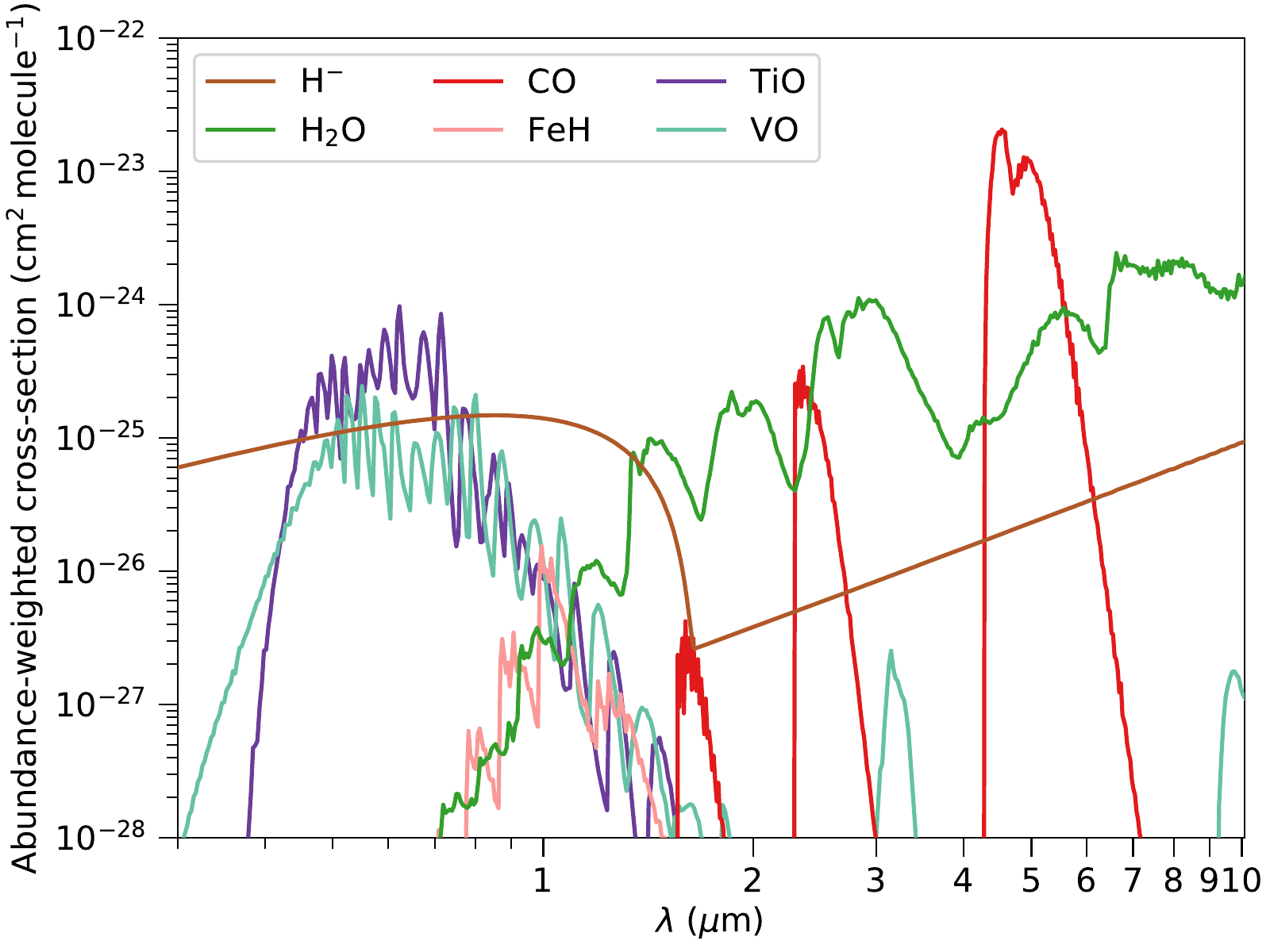}
\caption{Absorption cross-sections of important species for temperature 2700\,K and pressure 10\,mbar. Cross-sections have been weighted by the best-fit abundances at 10\,mbar, which approximately coincides with the near-infrared photosphere (see Figure \ref{fig:vertresults}).}
\label{fig:opacities}
\end{figure}

The contribution of different species to the atmospheric opacity is illustrated in Figure \ref{fig:opacities}. It shows absorption cross-sections for radiatively active species weighted by the corresponding abundances indicated in Figure \ref{fig:vertresults} at a pressure of 10\,mbar, coincident with the near-infrared photosphere. According to the model, the primary opacity source is H$^{-}$ across the G102 and $z^\prime$ passbands, as well as the short-wavelength half of the G141 passband. At longer wavelengths, H$_2$O dominates across the remainder of the G141 passband, as well as the $K_s$ and IRAC $3.6\um$ passbands, while CO dominates within the IRAC $4.5\um$ passband.

Although we do not detect any spectral features due to TiO or VO in our emission data, the presence of the muted H$_2$O band in the G141 passband could be hinting at the presence of these strong optical absorbers. Both \cite{2018ApJ...866...27L} and \cite{2018A&A...617A.110P} find that when TiO and VO are excluded as opacity sources in models of ultrahot Jupiter atmospheres, the $1.4\,\um$ H$_2$O band is entirely absent from the emission spectrum. This is because TiO and VO absorb a significant amount of incident stellar radiation at pressure levels above the near-infrared photosphere, even when their abundances have been depleted by thermal dissociation. When TiO and VO are removed as opacity sources, stellar radiation is able to penetrate deeper into the atmosphere, raising temperatures at the near-infrared photosphere by over 100\,K \citep{2018ApJ...866...27L}. This in turn should increase the thermal dissociation rates for H$_2$O enough to completely nullify the $1.4\,\um$ spectral band. Therefore, the fact that we observe a muted $1.4\,\um$ H$_2$O emission band is consistent with the presence of one or more optical absorbers, such as TiO and VO. 

Direct observational confirmation of the optical absorber/s responsible for the thermal inversion on the dayside hemisphere of WASP-121b will be challenging. Evidence for significant optical absorption at the day-night limb has been uncovered in the transmission spectrum, which may be due to VO, although no strong evidence for TiO has been found \citep{2018AJ....156..283E}. However, even if one or both of these species are present in significant quantities on the dayside, their broad emission bandheads are likely weakened by thermal dissociation and overlapping H$^-$ continuum opacity. One possible workaround could be to use high resolution spectroscopy with large aperture ground-based telescopes to doppler-resolve the narrow cores of the strongest TiO and VO lines just prior to eclipse, as these will be less affected \citep[e.g.][]{2017AJ....154..221N}. Using Figure \ref{fig:opacities} as a guide, TiO could supersede H$^{-}$ as the dominant opacity source regulating the emission for wavelengths shortward of $\sim 0.7\um$. Measurements made by the Transiting Exoplanet Survey Satellite (TESS) -- which observed WASP-121 throughout January 2019 -- will also help inform this picture. For example, if TiO and/or VO are present, a deeper eclipse depth would be expected within the TESS passband compared to the G102 passband, as the former extends across the $\sim 0.6$-$0.95\um$ wavelength range. Looking further ahead to the \textit{James Webb Space Telescope}, the second order of the Near Infrared Imager and Slitless Spectrograph (NIRISS) single-object spectrograph (SOSS) provides wavelength coverage across $\sim 0.6$-$0.8\,\um$, which encompasses a number of significant TiO and VO bands (Figure \ref{fig:opacities}). At even shorter wavelengths, the high resolution spectroscopy approach could in principle be used to doppler-resolve emission lines due to metals such as iron and titanium \citep{2018Natur.560..453H} or photochemical products such as SH \citep{2009ApJ...701L..20Z,2018AJ....156..283E}.  Although these latter species are not included in the model shown in Figure \ref{fig:opacities}, they have strong absorption lines at wavelengths shortward of $\sim 0.5\,\um$ and could potentially heat the upper atmosphere enough to produce the observed thermal inversion \citep{2018ApJ...866...27L}.

Finally, we note that the G102 data presented here provide the most direct evidence yet for H$^-$ emission in an exoplanet atmosphere. This is due to the clear departure from a blackbody at wavelengths shortward of $1.1\,\um$ (Figure \ref{fig:emspec_all_lin}), whereas previous claims have instead relied on model-dependent interpretations of G141 spectra that are indistinguishable from blackbodies \citep{2018ApJ...855L..30A,2018AJ....156...17K,2018AJ....156...10M}. The extensive ionization implied by this result for pressures below $\sim 100$\,mbar (Figure \ref{fig:vertresults}) could have important implications for atmospheric dynamics and energy transfer, including increased day-night heat redistribution due to H$_2$ recombination on the nightside hemisphere \citep{2018ApJ...857L..20B,2018RNAAS...2b..36K} and increased magnetic drag due to Lorentz forces \citep{2018ApJ...866...27L}. As \cite{2018AJ....156...17K} have shown for WASP-103b --- an ultrahot Jupiter that is in many respects similar to WASP-121b --- spectroscopic phase curves offer the most promising means of further constraining these fundamentally 3D phenomena.

\section{Conclusion} \label{sec:conclusion}

We have presented new secondary eclipse observations for the ultrahot Jupiter WASP-121b acquired with the G102 grism of HST/WFC3. These data extend the wavelength coverage of the measured emission spectrum from $1.1\,\um$ down to $0.8\,\um$. We performed a retrieval analysis of the combined emission dataset, improving upon our previous efforts by incorporating the effects of thermal dissociation and ionization. We confirm the detection of a thermal inversion and our best-fit model indicates that the temperature profile increases from approximately $2500$\,K to $2800$\,K across the $\sim 30$\,mbar to $5$\,mbar pressure range. The spectrum is well explained by H$^-$ emission for wavelengths shortward of $1.1\,\um$, a muted H$_2$O emission band at $1.4\,\um$, and overlapping CO and H$_2$O emission bands within the $4.5\,\um$ channel.  Under the assumption of chemical equilibrium, we find the dayside atmospheric metallicity is likely enriched by at least a factor of a few relative to solar and uncover no evidence for anomalous carbon and oxygen abundances.

\section*{Acknowledgements}

The authors are grateful to the anonymous referee for constructive feedback that improved the quality of this manuscript. Support for program GO-15135 was provided by NASA through a grant from the Space Telescope Science Institute, which is operated by the Association of Universities for Research in Astronomy, Inc., under NASA contract NAS 5-26555. JMG acknowledges funding from a Leverhulme Trust Research Project Grant and University of Exeter PhD Studentship. ALC is funded by a UK Science and Technology Facilities Council (STFC) studentship. BD acknowledges support from an STFC Consolidated Grant (ST/R000395/1). PT acknowledges support by the European Research Council under Grant Agreement ATMO 757858.

%%%%%%%%%%%%%%%%%%%%%%%%%%%%%%%%%%%%%%%%%%%%%%%%%%



%\bibliographystyle{mnras}
\bibliographystyle{apj}
\bibliography{wasp121}

%\begin{thebibliography}{99}
%\end{thebibliography}

%%%%%%%%%%%%%%%%%%%%%%%%%%%%%%%%%%%%%%%%%%%%%%%%%%

%%%%%%%%%%%%%%%%% APPENDICES %%%%%%%%%%%%%%%%%%%%%

%\appendix



%%%%%%%%%%%%%%%%%%%%%%%%%%%%%%%%%%%%%%%%%%%%%%%%%%

% Don't change these lines
\bsp	% typesetting comment
\label{lastpage}
\end{document}